\documentclass[10pt]{article}
\usepackage{amssymb,latexsym,amsmath,graphicx,subfigure}
\usepackage{amscd}
\usepackage{cite,psfig}

\newtheorem{thm}{\bf Theorem} %
\newtheorem{defn}[thm]{\bf Definition} %
\newtheorem{lem}[thm]{\bf Lemma} %
\newtheorem{prop}[thm]{\bf Proposition} %
\newtheorem{rem}[thm]{\bf Remark} %
\newtheorem{conj}[thm]{\bf Conjecture} %
\newcommand{\qed}{{$\square$}}

\begin{document}

\title{Dynamical systems on infinitely sheeted Riemann surfaces}

\author{Yuri N. Fedorov and David G\'omez-Ullate\\
 Department de Matem\`atica Aplicada I,\\
Universitat Polit\`ecnica de Catalunya,\\ Barcelona, E-08028
Spain.}
\maketitle
\begin{abstract}

This paper is part of a program that aims to understand the
connection between the emergence of chaotic behaviour in dynamical
systems in relation with the multi-valuedness of the solutions as
functions of complex time $\tau$. In this work we consider a
family of systems whose solutions can be expressed as the
inversion of a single
 hyperelliptic integral. The associated Riemann surface ${\mathcal R}\to {\mathbb
C}=\{\tau \}$ is known to be an infinitely sheeted covering of the
complex time plane, ramified at an infinite set of points whose
projection in the $\tau$-plane is dense. The main novelty of this
paper is that the geometrical structure of these infinitely
sheeted Riemann surfaces is described in great detail, which
allows to study global properties of the flow such as asymptotic
behaviour of the solutions, periodic orbits and their stability or
sensitive dependence on initial conditions. The results are then
compared with a numerical integration of the equations of motion.
Following the recent approach of Calogero, the real time
trajectories of the system are given by paths on $\mathcal R$ that
are projected to a circle on the complex plane $\tau$. Due to the
branching of $\mathcal R$, the solutions may have different
periods or may not be periodic at all.

%
%

\end{abstract}
\vskip 0.3cm \noindent {\bf Keywords:} Complex dynamics, Riemann
surface, inversion of hyperelliptic integrals, isochronicity,
sensitive dependence.


\newpage
\section{Introduction}

It has now become a classical subject to study the connections
between the integrability of a dynamical system (or the absence of
it) and the singularity structure and multi-valuedness of its
solutions. The discovery of this relation dates back to the work
of Painlev\'e and his collaborators \cite{painleve,gambier}, who
classified all second order ODEs whose solutions have movable
poles as their only singularities (now known as the Painlev\'e
property) and are therefore single-valued functions. Then
Kowalewskaya found a new integrable case of the heavy top by
 requiring that the solutions have the Painlev\'e property
\cite{kow1,kow2}. A number of techniques collectively known as
Painlev\'e-Kowalewskaya analysis have been developed over the last
thirty years (for a review see, for instance, \cite{RGB89}) to
test for this property of the solutions, essentially by seeking
for a formal solution near a singularity in terms of a Laurent
series, introducing it in the equations and determining the
leading orders and resonances (terms in the expansion at which
arbitrary constants appear). Painlev\'e analysis has been extended
to test for the presence algebraic branching (weak Painlev\'e
property,\cite{RGB89,RaDorGram}) by considering a Puiseaux series
instead of a Laurent series. These analytic techniques (which have
been algorithmized and are now available in computer packages)
constitute a useful tool in the investigation of integrability: in
many non-linear systems where no solution in closed form is known,
Painlev\'e analysis provides information on the type of branching
the general solution or special families might have. It has also
proved to be useful to identify special values of the parameters
for which  generically chaotic systems such as H\'enon-Heiles or
Lorenz  become integrable.

It is also natural to investigate the singularity structure of
solutions of chaotic dynamical systems. Tabor and his
collaborators initiated this study in the early eighties for the
Lorenz system \cite{WT81} and the Henon-Heiles
Hamiltonian\cite{CTW82}. They realized that the singularities of
the solutions in complex-time are important for the real-time
evolution of the system. The complex time analytic structure was
studied by extensions of the Painlev\'e analysis involving the
introduction of logarithmic terms in the expansion
--- the so called $\Psi$-series --- which provides a local
representation of the solutions in the neighbourhood of a
singularity in the chaotic regime. Their local analytic approach
was complemented by numerical techniques developed for finding the
location of the singularities in complex time and determining the
order of  branching \cite{corliss}. In all the chaotic systems
under study, they observed numerically that the singularities in
complex time cluster on a natural boundary with self-similar
structure \cite{CGTW83}. An analytic argument to explain the
mechanism that leads to recursive singularity clustering was given
in \cite{LT88}. Similar studies relating singularity structure,
chaos and integrability have been performed by Bountis and his
collaborators. Going beyond the local techniques described above,
the emphasis  was put on a global property of the complex
solutions: whether their Riemann surface has a finite or an
infinite number of sheets.  The authors of \cite{bount} proposed
to  classify the system as {\em integrable} in the first case and
as {\em non-integrable} for the second case.


These ideas actually go back to the work of Kruskal and his
poly-Painlev\'e test to detect non-integrability \cite{kruskal}.
Combining analytical and numerical results for a simple ODE,
Bountis and Fokas \cite{bountis_fokas} identified a chaotic
behaviour with the property that the singularities or branch
points of their solutions are dense. In their work, however, it is
not too clear whether the criterion is that such points should be
dense on the Riemann surface itself or that the projection of the
branch points is dense on the complex plane. The second condition
is obviously much weaker.

 In fact, as was indicated in \cite{AF, AMB}, there exists a
wide class of completely integrable Hamiltonian systems whose
solutions live on an infinitely sheeted covering of the complex
time plane with an infinite number of branch points that are
projected onto a dense set on the complex plane and yet are
isolated on the covering surface.


 Painlev\'e analysis is a local technique and does not provide
information on the {\em global} properties of the Riemann surfaces
of the solutions (the number and location of the movable branch
points, how the different sheets of the Riemann surface are
connected together, etc.), which can be important for
understanding of the dynamics of the system.

Aware of this limitation, F. Calogero initiated a new line of
research by studying the real time dynamics of a system as a path
on the Riemann surface of its solution in complex time. In
particular, Calogero and Fran{\c{c}}oise have shown that many
isochronous systems can be written by a suitable modification of a
large class of complex ODEs \cite{CF02}. Using local analysis and
numerical integration in two many-body systems in the plane
\cite{CFS03,CS02}, it was discovered that the periodic solutions
could have a very high period, and the mechanism for the period
increase was conjectured. However, in those systems it was not
possible to achieve a complete description of the Riemann surface.
Although it was proved that all singularities are finite order
branch points, the presence of an infinite number of them could
still produce aperiodic motion as the solution visits an infinite
number of sheets. In the subsequent work \cite{CGSS}, the Riemann
surface of the solution was infinitely sheeted for generic values
of the parameters, yet it could be described in full detail. Using
these rather novel techniques, Calogero et al. were able to derive
analytic expressions for the periods and prove sensitive
dependence on initial conditions. In \cite{CGSS} the projection of
square-root branch points of the infinitely sheeted Riemann
surface cover densely a circle on the complex plane for generic
values of the parameters. As a further step on the way to
understand the connection between chaotic behaviour and analytic
structure, it was natural to extend the study to Riemann surfaces
whose branch points projected onto $\mathbb C$ do not cover
densely a one-dimensional curve in $\mathbb C$, but the whole
complex plane. The candidate objects for this study are the
Riemann surfaces associated to the inversion of hyper-elliptic
integrals, first described by Jacobi \cite{Mark}.


This is precisely the case of the dynamical systems treated in
this paper. The text is organized as follows: in Section
\ref{sec:dynamic} we introduce a family of simple,  generally
nonautonomous, dynamical systems whose solutions can be understood
as lifting a circular path to a Riemann surface associated to the
inversion of a hyperelliptic integral.  Global geometrical
properties of such surfaces  for different cases are described in
Section \ref{sec:inversion}. The algorithm that performs the lift
of the circular contour on the Riemann surface and possible
behaviours of the corresponding solutions are described in Section
\ref{sec:circular}. Here we make a conjecture that all the
solutions are either periodic or singular (i.e., they escape to
infinity in finite time).

The trajectories of the system for different initial conditions
are shown in Section \ref{sec:numerical}, where the path on the
Riemann surface is compared with a numerical integration of the
equations of motion. Note that trajectories of complex
Hamiltonians have been also studied  numerically by Bender et al.
in \cite{bender}. Although the systems they study are different
from the ones treated here, their approach is related to the one
we propose and the observed complexity of the orbits is similar to
the results of Section \ref{sec:numerical}. Finally, conclusions
and directions of future work are outlined in Section
\ref{sec:conclusions}.

\section{Two classes of dynamical systems whose solutions travel on the
Riemann surface describing inversion of a hyperelliptic
integral}\label{sec:dynamic}

In \cite{fla88} Flashka drew attention to a simple class of
integrable Hamiltonian systems in ${\mathbb R}^4=(x, y, p_x,p_y)$
with real time $t$, whose Hamiltonians and additional commuting
integrals are given by the real and imaginary parts of arbitrary
complex polynomials $H(\zeta,\mu)$, $\zeta=x+i y$, $\mu=p_x+i
p_y$. In contrast to the case of the Liouville integrability, the
generic real level surfaces
$$
{\mathcal I}_E = \{x,y,p_x,p_y\mid H(\zeta,\mu)=E, E\in {\mathbb
C}\}
$$
are not 2-dimensional tori or products of lines and circles, but
algebraic Riemann surfaces with some points removed. Then, instead
of straight line motion on the tori, one has flows on such
surfaces, which may have a rather complicated behaviour. Solutions
of such systems are given by inversion of Abelian integrals on
${\mathcal I}_E$ and posesses the weak Painlev\'e property.

Motivated by problems of mechanics and mathematical physics, we
restrict ourselves to complex Hamiltonians
\begin{equation}\label{eq:E}
H =\frac{1}2 (\zeta')^2+V(\zeta) ,
\end{equation}
which generate second order complex ODE
\begin{equation} \label{newton}
\zeta''= -\frac{d V(\zeta)}{d\zeta}
\end{equation}
where $\zeta=\zeta(\tau)$, both $\zeta$ and $\tau$ are complex and
the prime denotes differentiation of $\zeta$ with respect to
$\tau$.

By fixing the energy, $H=E$, the complex solution is formally
obtained as the inversion of the following quadrature
\begin{equation}\label{eq:intV}
\tau-\tau_0 =\int_{\zeta_0}^\zeta
\frac{d\eta}{\sqrt{2(E-V(\eta))}},
\end{equation}
where $\tau_0$ is the initial time and $\zeta_0= \zeta(\tau_0)$.
Whenever the potential $V(\zeta)$ is a polynomial or a rational
function of $\zeta$, the above integral is in general
hyperelliptic, and $\zeta(\tau)$ is a multi-valued or even
infitely-valued function on the complex plane $\tau$.

We emphasize that a single hyperelliptic integral often appears as
a quadrature in various integrable problems of mechanics and
mathematical physics, and the last decade the problem of its
inversion revokes increasing attention, \cite{AF,Alber,ERP,Van}.


On the other hand, by introducing real time $t$ such that
$\tau=\varkappa t+\tau_0$, $\varkappa$ being arbitrary nonzero
complex constant, the real and imaginary parts of $\zeta(\tau(t))$
give a solution of the Hamiltonian system in ${\mathbb R}^4=(x, y,
p_x,p_y)$ with a real Hamiltonian proportional to $\mbox{Re}
\,\varkappa \mbox{Re} H + \mbox{Im} \varkappa \,\mbox{Im} H$ and
describe flows on the hyperelliptic curve $H(\zeta,\mu)=E$.

Some properties of such flows were studied in \cite{bates} and,
from the algebraic topology point of view, in \cite{novikov} (see
also references therein).

In our paper we focus on a similar yet different class of systems
that describe not straight-line, but circular motion on the
complex plane $\tau$.

\subsubsection*{Calogero's circular motion}

F. Calogero has shown that a large class of complex ODEs can be
modified in such a way that the resulting system  posseses a large
number of periodic solutions \cite{Cal02}. Following Calogero's
approach, let us construct another class of systems associated to
the quadrature (\ref{eq:intV}) by performing the following change
of independent and dependent variables,
\begin{eqnarray}\label{eq:trick1}
\tau -\tau_0 &=&\frac{1}{{\rm i} \omega}\, \left({\rm e}^{{\rm i}\omega t}-1\right),\\
\zeta(\tau)&=&{\rm e}^{{\rm i}\alpha\omega t}\,z(t),
\label{eq:trick2}
\end{eqnarray}
where $\omega>0$ is a real parameter that plays the role of a
frequency to which the fundamental period $T= 2 \pi/\omega$ is
associated and $\alpha\in \mathbb R$ is a parameter still to be
determined. The effect of the transformation \eqref{eq:trick2} is
to introduce the {\em real} time evolution as a parametrization of
a circle in the complex $\tau$-plane. Thus the name {\em circular
motion} as opposed to the {\em straight line} motion studied by
Flaschka. If the solutions of the complex ODE were analytic or
meromorphic functions of $\tau$, then the resulting real time
evolution would be periodic, but if the solutions are multi-valued
functions of $\tau$ (possibly infinitely-valued) more complex
behaviours are possible.

In what follows, for the sake of simplicity, we concentrate on the
case
\[V(\zeta)=\zeta^{k+1},\qquad k\in\mathbb Z^+ . \]
The restriction of $V(\zeta)$ to a single power rather than an
arbitrary polynomial is by no means essential for the analysis of
the geometrical properties of the Riemann surface performed in
Section \ref{sec:inversion}. As a matter of fact, the results of
this Section are valid for an arbitrary polynomial function
$V(\zeta)$.

Under the transformation \eqref{eq:trick1}--\eqref{eq:trick2}, the complex ODE (\ref{newton}) with potential $V(\zeta)=\zeta^{k+1}$ 
transforms into
\begin{equation}\label{eq:z2}
{\rm e}^{{\rm i}\omega (\alpha-2) t}\left[ \ddot z+ {\rm i} \omega
(2\alpha-1) \dot z + \alpha(1-\alpha) \omega^2\,z
\right]=-(k+1){\rm e}^{{\rm i}\alpha k\omega t}\,z^k.
\end{equation}
If we require the transformed system to be {\em autonomous} then,
necessarily,
\begin{equation} \label{al}
\alpha = \frac{2}{1-k}, \qquad k\in{\mathbb Z}^+ .
\end{equation}
In this case we obtain the following second order non-linear ODE
for the function $z=z(t)$:
\begin{equation}\label{eq:z}
\ddot z+ {\rm i} \left( \frac{3+k}{1-k}\right)\,\omega\dot z-
 \frac{2+2k}{(1-k)^2}\,\omega^2 z= -(k+1) z^k,
\end{equation}
where  $z\in\mathbb C$ but now $t\in \mathbb R$ is an ordinary
(real) time variable, and dots denote derivative of $z(t)$ with
respect to its argument. The complex variable
$\zeta(t)\equiv\zeta(\tau(t))$ evolves in time according to the
following non-autonomous ODE:
\begin{equation}\label{eq:zeta(t)}
\ddot\zeta -{\rm i} \omega \dot \zeta= -(k+1)\, {\rm e}^{2 {\rm i}
\omega t}\,\zeta^k,
\end{equation}
which corresponds to setting $\alpha=0$ in \eqref{eq:z2}. The
solutions of \eqref{eq:zeta(t)} and \eqref{eq:z} are of course
related by the simple transformation \eqref{eq:trick1}, a change
into a rotating frame of reference with constant angular velocity,
and the relation between the initial data is:
\begin{equation}\label{initialdata}
\zeta(0)=z(0),\qquad \zeta'(0)=\dot z(0) + {\rm i} \alpha \omega
z(0).
\end{equation}

 If the potential $V(\zeta)$ were not a
simple power $\zeta^{k+1}$ but a polynomial in $\zeta$, Calogero's
trick would lead to a system with non-autonomous $t$-periodic
terms which can be interpreted as external periodic forcing. The
geometrical description of the Riemann surface associated to the
inversion of \eqref{eq:intV} when $V(\zeta)$ is an arbitrary
polynomial of degree $5$ is given in the following section.

\subsubsection*{Three different examples.}

We will choose three different integer values of $k$ in
(\ref{eq:z}) which lead to systems with quite different
properties.

 \begin{enumerate}
\item If we set $k=2$ in (\ref{eq:z}) and write it out in terms of
its real and imaginary components $z=x+{\rm i} y$, we obtain the
following system of two coupled second order real ODEs:
\begin{equation}\label{eq:xy}
\begin{aligned}
\ddot x +5 \omega\dot y- 6\omega^2 x &= -3 (x^2-y^2)\,,\\
\ddot y -5 \omega\dot x- 6\omega^2 y&= -6xy.
\end{aligned}
\end{equation}
This system has one surprising property: it is {\it isochronous},
i.e., for any set of initial data $\{x(0),y(0),\dot x(0),\dot
y(0)\}$ the solution $\{x(t),y(t)\}$ is a periodic function of
time with period $T=\frac{2\pi}{\omega}$. The remarkable fact is
that isochronicity is compatible with the non-linear character of
(\ref{eq:xy}).

Indeed, as follows from the quadrature (\ref{eq:intV}) for $k=2$,
its inversion $\zeta(\tau)$ is an elliptic function of $\tau$,
therefore meromorphic and single-valued. Since $\tau(t)$ is
periodic in $t$ with period $T=2 \pi/\omega$, it follows from
(\ref{eq:trick2}), (\ref{al}) that the solution $z(t) = {\rm e}^{2
{\rm i} \omega t}\,\zeta(\tau(t))$ is also periodic with the same
period.
\vskip 0.2cm

 \item In the second example, we set $k=4$ in (\ref{eq:z})
which leads to the system
\begin{equation}\label{eq:xyk4}
\begin{aligned}
\ddot x +(7 \omega/3) \dot y- (10\omega^2/9) x&= -5\, (x^4+y^4-6x^2y^2)\,,\\
\ddot y -(7 \omega/3)\dot x- (10\omega^2/9) y&= -20\,xy(x^2-y^2).
\end{aligned}
\end{equation}
In this case it is no longer true that all the solutions have the same period, moreover, they may be
not periodic at all, since the function $\zeta(\tau)$ obtained by inverting the
hyperelliptic integral
\begin{equation}\label{int5}
\tau-\tau_0 =\int_{\zeta_0}^\zeta \frac{d\eta}{
\sqrt{2(E-\eta^5})},
\end{equation}
is no longer single-valued. In fact (see Section
\ref{sec:inversion}) it can be considered as a function on an {\it
infinitely} sheeted ramified covering $\mathcal R\to {\mathbb
C}=\{\tau\}$ such that the projections of the branch points at the
complex plane form a dense set \cite{Mark}. This example motivates
the study of geometric properties of the Riemann surface $\mathcal
R$ and its covering of $\mathbb C$. As we shall see, the evolution
of the dynamical system (\ref{eq:xyk4}) can be analyzed by
following a circular contour $\tau(t)$ lifted to the infinitely
sheeted covering $\mathcal R$. \vskip 0.2cm \item For the third
example we choose $k=5$ in (\ref{eq:z}). The equations defining
the dynamical system are now
\begin{equation}\label{eq:xyk5}
\begin{aligned}
\ddot x +2\, \omega\dot y-3/4 \,\omega^2 x&= -6x(x^4+5 y^4-10 x^2 y^2)\,,\\
\ddot y -2\, \omega\dot x- 3/4\,\omega^2 y&= -6y(y^4+5x^4-10
x^2y^2).
\end{aligned}
\end{equation}
and the function $\zeta(\tau)$ is obtained by inverting the
integral
\begin{equation}
\tau-\tau_0 =\int_{\zeta_0}^\zeta
\frac{d\eta}{\sqrt{2(E-\eta^6)}}\,.
\end{equation}
Despite the similarity between this case and the preceding one,
the behaviour of the solutions is very different. As we shall see
in Remark 3.1 below, for any initial condition the solution of
(\ref{eq:xyk5}) is periodic with period either $T$ or $2T$.

\end{enumerate}

The above examples illustrate the importance of studying the
detailed structure of the Riemann surface that describes the
inversion of a hyperelliptic integral.

\section{Inversion of a single hyperelliptic integral and  the associated Riemann
surface}\label{sec:inversion}

 We first recall some necessary basic
facts of the theory of algebraic Riemann surfaces. Let
$\eta_b^{(1)},\dots,\eta_b^{(5)}$ be arbitrary distinct complex
numbers. The genus 2 surface
\begin{equation} \label{gam}
\Gamma=\{(\eta,\mu)|\, \mu^2=P_5(\eta) \}, \qquad P_5(\eta)= -
(\eta-\eta_b^{(1)}) \cdots (\eta-\eta_b^{(5)}),
\end{equation}
can be represented as 2-fold covering of the Riemann sphere $\bar
{\mathbb C} =\{ \eta\}\cup \infty$ ramified at six branch points
$$
Q_1=(\eta_b^{(1)},0), \quad \dots, \quad Q_5=(\eta_b^{(5)},0),
\quad \infty .
$$
Points on different $\bar{\mathbb C}$-sheets that have the same $\eta$-coordinate have
$\mu$-coordinates with opposite signs.
For this reason we denote the sheets as $\bar {\mathbb C}^+$ (upper sheet) and
$\bar{\mathbb C}^-$ (lower sheet).
\begin{figure}[h,t]\label{cyc.fig}
\begin{center} 
\includegraphics[width=.6\textwidth]{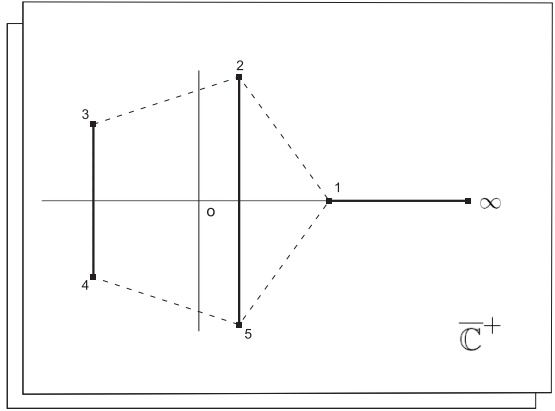}
\end{center} \caption{ The two sheets $\bar {\mathbb C}^\pm$ of $\Gamma_5$ with the branch points and cuts (solid lines). }
\end{figure}
The passage from one $\bar{\mathbb C}$-sheet to the other is
realized through any of the three cuts that join the pairs $(Q_2,
Q_5)$, $(Q_3, Q_4)$ and $(Q_1, \infty)$ on $\bar {\mathbb C}^+$
and $\bar {\mathbb C}^-$ and that do not intersect each other. Let
us fix a canonical basis of cycles $a_1,a_2,b_1, b_2$ on $\Gamma$
such that
 $$
{a}_{i}\circ {a}_{j}={b}_{i}\circ {b}_{j}=0, \quad a_{i}\circ b_{j}=\delta_{ij}, \qquad
i,j=1,2,
$$
where $\gamma_{1}\circ \gamma_{2}$ denotes the intersection index
of the cycles $\gamma_{1},\gamma_{2}$. Following the common
convention, $a$-cycles lie entirely on the upper sheet
$\bar{\mathbb C}^+$ and embrace one of the cuts while $b$-cycles
run through both sheets. We assume that $a_1$ embraces $(Q_1,
\infty)$ and $b_1$ joins $\infty$ with $Q_3$, so that the
projections of cycles $a_1, b_1$ embrace $\infty$. For the
pentagonal curve $\Gamma_5= \{ (\eta,\mu)|\,\mu^2=\eta^5-E\}$, the
$\eta$-coordinates of the finite branch points are
\[ \eta_b^{(k)}=\exp\left(\frac {2\pi {\rm i}(k-1)}{5}  +\frac 15
\arg E\right)\sqrt[5]{|E|}, \quad k=1,\dots,5.\] For real $E>0$,
the branch points and cuts are depicted in Figure 1, where $\bar
{\mathbb C}^\pm$ are shown as planes and $Q_j$ are denoted just by
their indices $j$. Figure 2 also shows the canonical cycles, where
segments of the cycles $b_1, b_2$ on the upper and lower sheets
are given by solid and dashed lines respectively. Since the branch
points $(Q_i,0)$, $i=1,\dots,5$ are just the five complex roots
$E^{1/5}$, for generic (complex) values of $E$ their position is
simply obtained by a rotation and scaling of the pentagon of
Figure 2.
\begin{figure}[h,t]\label{cycles.fig}
\begin{center} 
\includegraphics[width=.6\textwidth]{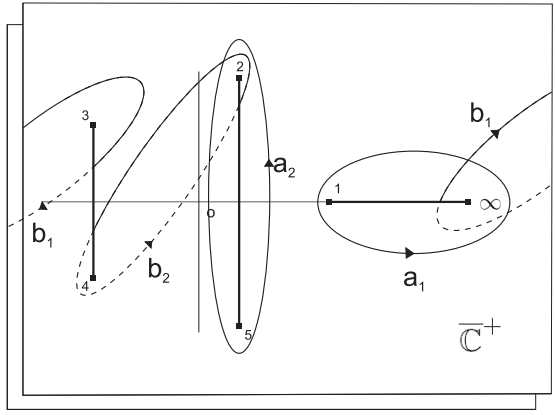}
\caption{ Canonical cycles $a_i$ and $b_i$, $i=1,2$ for the
pentagonal curve $\Gamma_5= \{ (\eta,\mu)|\,\mu^2=\eta^5-E\}$.
Segments of cycles on the upper and lower sheet are depicted as
solid and dashed lines respectively. }
\end{center}
\end{figure}

We introduce next the following basis of holomorphic differentials
$\left(\frac {d\eta} \mu ,  \frac{\eta\,d\eta} \mu \right)$ on
$\Gamma$ and
consider in detail the problem of inverting the hyperelliptic integral 
\begin{equation}  \label{int_main}
\tau= \int_\infty^P  \frac {d\eta} {\sqrt{2}\, \mu} , \qquad
P=(\eta,\mu) \in \Gamma .
\end{equation}
The integral has four independent complex periods defined by:
\begin{equation}\label{periodvecs}
V_1= \oint_{a_1}\frac {d\eta} {\sqrt{2} \,\mu} , \quad V_2=
\oint_{a_2}\frac {d\eta} {\sqrt{2}\, \mu} ,  \quad V_3=
\oint_{b_1}\frac {d\eta} {\sqrt{2}\, \mu} , \quad V_4=
\oint_{b_2}\frac {d\eta} {\sqrt{2}\, \mu} .
\end{equation}
It follows that the inverse complex functions $\eta(\tau)$,
$\mu(\tau)$ should have four generally independent periods on
$\mathbb C$. This is not possible, a fact that was already noted
by Jacobi (see \cite{Mark} for an historical account). If the
periods $V_1,\dots,V_4$ are not commensurable, then the functions
$\eta(\tau)$, $\mu(\tau)$ cannot be singled valued on ${\mathbb
C}=\{\tau\}$, but they can be regarded as single-valued on an
infinitely sheeted covering ${\mathcal R}\mapsto {\mathbb C}$.
This Riemann surface has an infinite number of ramification points
whose projections onto the $\tau$-plane form a dense set.

\begin{rem}\label{rem:bolza}
According to the Weierstrass--Poincar\'e theory of reductions of
Abelian functions, (see e.g. the surveys \cite{be02, BEBIM}), the
periods of ${d\eta}/\mu$  may become commensurable in special
cases when the genus 2 curve $\Gamma$ covers an elliptic curve
$\mathcal E$ and the differential $d\eta /\mu$ is a pull-back of a
holomorphic differential on $\mathcal E$. Then ${\mathcal
R}\mapsto {\mathbb C}$ is just a finite order covering. For genus
2 curves admitting a finite order automorphism, a complete list of
their Riemann matrices that characterize such coverings was given
by Bolza in \cite{Bolza} (see also \cite{LB}).

We note that the pentagonal curve $\Gamma_5= \{ \mu^2=E-\eta^5\} $
 does not belong to this list,
hence its periods are incommensurable (see also Proposition
\ref{periods-5} and relation (\ref{coll}) below). On the contrary,
the sextic curve $\Gamma_6 = \{ \mu^2=E-\eta^6 \}$ does appear in
Bolza's list: it is
a 2-fold ramified cover of two 
elliptic curves,
\begin{gather*}
\begin{CD}
{\mathcal E}_1 @ < \pi_1 << \Gamma_6 @ > \pi_2 >> {\mathcal E}_2, \end{CD} \\
 {\mathcal E}_1=\{w_1^2=E-z_1^3 \}, \quad {\mathcal E}_2=\{w_2^2=z_2(E-z_2^3)\}, \\
(z_1,w_1)= \pi_1(\eta,\mu)=(\eta^2, \mu), \quad  (z_2,w_2)=
\pi_2(\eta,\mu)=(\eta^2, \eta\mu).
\end{gather*}
Then $d\eta/\mu$ is the pullback of the holomorphic differential
$dz_2/w_2$ on ${\mathcal E}_2$. As a result, the corresponding
covering ${\mathcal R}\mapsto {\mathbb C}$ is only 2-fold, which
implies that, similarly to the case of the system (\ref{eq:xy}),
all the solutions of equations (\ref{eq:xyk5}) are periodic,
either with period $T=2\pi/\omega$ or $2T$.
\end{rem}

\begin{rem}
Note that for an arbitrary curve $\Gamma$ given by (\ref{gam}),
the differential $d\eta /\mu$ has a double zero at $\infty$ and no
zeros elsewhere. Namely, choosing a local coordinate $\delta(P)$
in a neighborhood of $\infty\in \Gamma$ such that
$\delta(\infty)=0$ and $\eta=1/\delta^2$, $\mu= O(1/\delta^5)$,
one has $d\eta /\mu= O(\delta^2)\, d\delta$ and, therefore,
$\tau=O(\delta^3)$, $\delta=O\left(\sqrt[3]{\tau}\right)$. This
implies that the functions $\eta(\tau),\mu(\tau)$ have only
critical points $\tau_*$ with the behavior
\begin{equation} \label{fraction}
\begin{gathered}
\eta(\tau)= (\tau-\tau_*)^{-2/3}(c_0 + c_1(\tau-\tau_*)^{2/3}+ c_2 (\tau-\tau_*)^{4/3}+ \cdots), \\
\mu(\tau)= (\tau-\tau_*)^{-5/3}(d_0 + d_1(\tau-\tau_*)^{2/3}+ d_2
(\tau-\tau_*)^{4/3}+ \cdots),
\end{gathered}
\end{equation}
where $c_j$ and $d_j$ are constants. Thus, the singularities
$\tau_*$ are $3^{\footnotesize\mbox{rd}}$ order branch points of
the covering ${\mathcal R}\mapsto {\mathbb C}$ and there is no
other type of branching.
\end{rem}
\medskip

Despite the complexity of the Riemann surface $\mathcal R$, it is
possible to give an explicit description of its geometric and
topological structure, which enables us to describe the behavior
of the solutions of system (\ref{eq:xyk4}). Such an explicit
description of $\mathcal R$ is new to our best knowledge.

\subsubsection*{Description of a sheet of $\mathcal R$} 
\begin{figure}[h,t]\label{Brule.fig}
\begin{center}
\includegraphics[width=.6\textwidth]{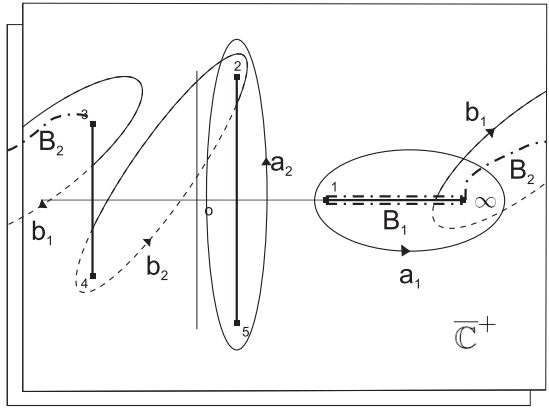}
\caption{ Canonical cycles $a_i$ and $b_i$ and the new cuts $B_1$
and $B_2$ (dashed-dotted lines) needed to define the B-rule.}
\end{center}
\end{figure}

We start by
making some preliminary observations. Namely, on the $\bar{\mathbb
C}^+$--sheet of $\Gamma$ we make two extra cuts $B_1^+$ and
$B_2^+$ that join $\infty$ with $Q_1$ and $Q_3$, respectively (see
Figure \ref{Brule.fig}). We assume that $B_1^+$ and $B_2^+$ are
embraced by projections of the cycles $a_1$ and $b_1$. It is
always possible to choose the cuts in such a way that path
integral (\ref{int_main}) along each of them yields a straight
line segment on the complex $\tau$-plane\footnote{Note that
generally the integration along a straight line segment on the
$\eta$-plane produces a curvilinear segment on the $\tau$-plane.
The only exception is the integration along a segment on the real
axis that does not contain the branch points of $\Gamma$.}. We
also introduce the two cuts $B_1^-$ and $B_2^-$ on the
$\bar{\mathbb C}^-$--sheet with the same properties (in Figure
\ref{Brule.fig}  such cuts are depicted as the bold dotted lines.)

\begin{defn}[B-rule]
Consider the hyperelliptic integral (\ref{int_main}) and allow the
point $P$ to range over the whole curve $\Gamma$ in such a way
that the integration path from $\infty$ to $P$ {\em does not cross
any of the cuts $B_1^\pm$ and  $B_2^\pm$}. We shall refer to this
condition as the {\em Border rule}, or for simplicity, the
$B$-rule.
\end{defn}

Note that since the cycles $a_1$ and $b_1$ cross the cuts $B_2$
and $B_1$ respectively, the corresponding periods $V_1$ and $V_3$
are not allowed by the $B$-rule, and we shall refer to them as the
{\em forbidden periods}. Consequently, we shall refer to the other
periods $V_2$ and $V_4$ as the {\em allowed periods}.

\begin{prop} \label{home_sheet}
Under the B-rule, the values of the integral $\tau(P)$ range over
the whole complex plane $\mathbb C$ except for an infinite number
of non-intersecting identical domains $W_{ij}$, $i,j\in {\mathbb
Z}$ which shall be called {\em windows}. Each window is a
parallelogram spanned by the forbidden periods $V_1$ and $V_3$,
while the window lattice $W_{ij}$ is generated by the allowed
periods $V_2$ and $V_4$. The corners of the windows correspond to
the image of the point at
 $\infty \in \Gamma$. At each pair of points $\tau_1, \tau_2$
lying on the opposite edges of $W_{ij}$ such that
$\tau_2=\tau_1+V_1$ or $\tau_2=\tau_1+V_3$ the coordinates
$\eta,\mu$ take the same values.
\end{prop}


In other words, the window $W_{i,j+1}$ ($W_{i+1,j}$ ) is obtained
from $W_{i, j}$ shifting by period $V_2$ ($V_4$). When $P\in
\Gamma$ makes a linear combination of cycles $a_2,b_2$, the image
$\tau(P)$ shifts by the corresponding combination of $V_2, V_4$.
More specifically, the two edges of the windows parallel to $V_1$
($V_3$) are the image of the two opposite sides of the cut
$B_1^\pm$ ($B_2^\pm$).

\begin{figure}[h,t]
\begin{center}
\includegraphics[width=.75\textwidth]{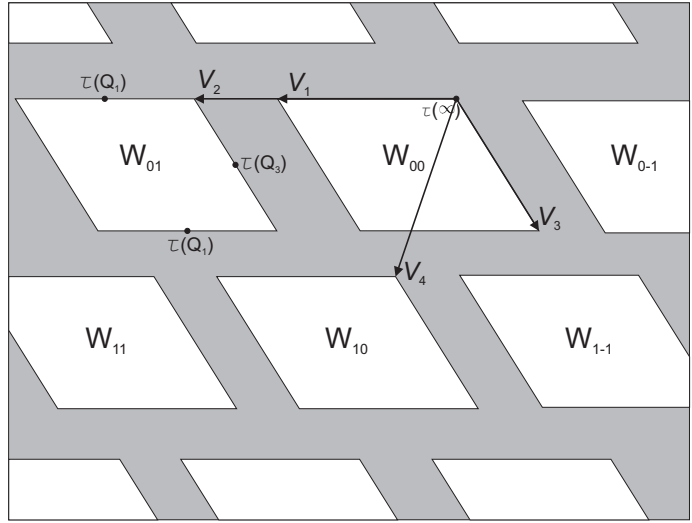}
\caption{ Domain $\mathcal{D}$ (shaded area) and windows $W_{ij}$.
We can consider this to be one of the infinite number of sheets of
the Riemann surface $\mathcal R$ \label{lattice.fig}}
\end{center}
\end{figure}

It follows that under the border rule, the two inverse functions
$\eta(\tau)$ and $\mu(\tau)$ are {\em quasi-elliptic}, in the
sense that they are doubly periodic with the allowed periods $V_2$
and $V_4$, but they are not defined in the whole complex
$\tau$-plane. The domain of definition ${\mathcal D}={\mathbb
C}\setminus \{\cup\; W_{ij}\}$ can be regarded as one of the
infinite set of sheets of the Riemann surface $\mathcal R$. Inside
each parallelogram of periods $V_2$ and $V_4$ the functions have 4
poles lying at the corners of a window $W_{ij}$. This is the
equivalent of a fundamental domain for the ``quasi-elliptic''
functions $\eta(\tau)$ and $\mu(\tau)$under the B-rule. For a general genus 2 curve $\Gamma$ with two real periods, 
the domain ${\mathcal D}$ is depicted in Figure \ref{lattice.fig} 
 as the grey shaded area.

\noindent\textbf{Proof of Proposition \ref{home_sheet}.} Let us
first study the integral $\tau(P)$ when $P\in \Gamma$ moves along
the cuts $B_1^\pm$ and $B_2^\pm$ without crossing them. When $P$
moves on $\bar{\mathbb C}^+$ from $\infty$ to $Q_1$ along the
upper edge of $B_1^+$ and returns to $\infty$ along the lower edge
of $B_1^+$, the point $\tau(P)$ performs a straight line motion
along the forbidden period vector $V_1$. If $P$ continues moving
on $\bar{\mathbb C}^+$ from $\infty$ to $Q_1$ along the lower edge
of $B_2^+$ and returns to $\infty$ along the upper edge of $B_2^-$
on $\bar{\mathbb C}^-$, the integral $\tau(P)$ traces the second
forbidden period vector $V_3$.  If $P$ traces the same two motions
now on the opposite $\bar{\mathbb C}$-sheets of $\Gamma$, the
point $\tau(P)$ moves subsequently along the vectors $-V_1$ and
$-V_3$. Corresponding to this sequence of motions of $\tau$ along
the cuts the integral $\tau(P)$ traces a parallelogram $W$ spanned
by the forbidden periods on the $\tau$-plane. The coordinates
$\eta,\mu$ of $P$ coincide along the points on the upper edge of
$B_1^+$ ($B_1^-$) identified with the the lower edge of $B_1^-$
($B_1^+$),  and they also coincide along the opposite points on
the upper and lower edges of $B_2^+$ ($B_2^-$). Under the B-rule
the point $\tau(P)$ cannot reach inside of a window $W$. Outside
of $W$, $\tau(P)$ behaves as an elliptic integral of the first
kind with  periods $V_2$ and $V_4$, therefore $\tau(P)$ cannot
reach inside of an infinite number of windows $W_{ij}$ obtained
from $W$ by translations by the allowed periods. Since the map
$\tau(P)$ is analytical, in ${\mathbb C}=\{\tau \}$ each $W_{ij}$
has a small neighborhood covered by the image of $P$, hence the
parallelograms cannot intersect. Finally, the proof that $\tau(P)$
reaches any point of ${\mathcal D}={\mathbb C}\setminus \{\cup\;
W_{ij}\}$ goes along the same lines as that for any elliptic
integral of the first kind. \qed

\subsection{Gluing  the ${\mathcal D}$-sheets together.}

Let us first introduce some notation for the different edges of a
window $W_{ij}$ on $\mathcal D$: we shall label them as $N$
(``North''), $W$ (``West''), $S$ (``South''), and $E$ (``East'').
By convention, the $N$- and $S$-edges are spanned by the period
$V_1$, while the $W$- and $E$-edges lie along $V_3$. Next, we
assume that the $B$-rule can be violated and we allow the
integration path from $\infty$ to $P$ to cross one of the cuts
$B_1^\pm, B_2^\pm$ only once. If the path crosses $B_1^-,\;
(B_1^+)$, then the point leaves the domain $\mathcal D$ through
$S$-edge (respectively, $N$-edge) of a window $W_{ij}$.

In view of Proposition \ref{home_sheet}, this motion can be viewed
as the passage to a different sheet ${\mathcal D}'$ of the surface
$\mathcal R$, which is a translate of an identical copy of
$\mathcal D$ and has the same window lattice structure. On
crossing the $S$-edge of $W_{ij}$ on $\mathcal D$, the image
$\tau(P)$ immediately arrives at the $N$-edge of a window
$W_{ij}'$ of ${\mathcal D}'$ and then continues travelling on this
new sheet. That is, the $N$-edge of a window $W_{ij}'$ on
${\mathcal D}'$ is glued to the $S$-edge of $W_{ij}$ on ${\mathcal
D}$ in such a way that the corners are connected to corners and at
the identified points the values of $\eta, \mu$ coincide. Next, on
crossing the $N$-edge of the window $W_{ij}$ on $\mathcal D$,
$\tau(P)$ enters another sheet $\mathcal D''$ from the $S$-edge of
a window $W_{ij}''$ on $\mathcal D''$ .
Similarly, when the integration path on $\Gamma$ crosses $B_2^-,\;
(B_2^+)$, then $\tau(P)$ leaves $\mathcal D$ through the $E$-
(respectively, $W$-) edge of the window $W_{ij}$ and arrives at
other sheets of $\mathcal R$ from the $W$- (respectively, $E$-)
edges of windows on these new sheets.

Since there is an infinite number of windows on $\mathcal D$, one
might think that each sheet is connected to an infinite set of
different $\mathcal D$-sheets of $\mathcal R$. However, the actual
structure of $\mathcal R$ is simpler and completely described by
the following short proposition.

\begin{prop} \label{whole_R} The Riemann surface $\mathcal R$ is a union of an
infinite number of identical $\mathcal D$-sheets glued to each
other along the corresponding opposite edges of their windows. Any
arbitrarily chosen ``home'' sheet ${\mathcal D}_H$ is connected to
precisely four other sheets ${\mathcal D}_{HN}, {\mathcal D}_{HE},
{\mathcal D}_{HS}, {\mathcal D}_{HW}$ and the passage from
${\mathcal D}_H$ to them is realized by crossing respectively the
$S$-, $W$-, $N$-, and, $E$-edge of {\em any window} $W_{ij}$ on
${\mathcal D}_H$.
\end{prop}

\noindent\textbf{Proof of Proposition.}
 We only need to prove that passage through $S-$ (or $W$- or
$N$- or $E$-) edges of different windows of ${\mathcal D}_H$ leads
to the same ${\mathcal D}$-sheet of ${\mathcal R}$.

Let $p (t)$ be a path on $\Gamma$ that starts at a point $P_0$, crosses the border
$B_1^\pm \cup B_2^\pm$ once and ends at a point $P_1$.
Next, let $ p^{(1)} (t) $ be the path 
$p(t)$ followed by a linear combination of cycles $a_2, b_2$ and
$p^{(2)} (t)$ be the same linear combination of cycles (with
starting point $P_0$) followed by the path $p(t)$. Then the
integrals $\tau(p^{(1)} (t))$, $\tau(p^{(2)}(t))$ give paths
$\gamma_1, \gamma_2$ on ${\mathcal R}$ that start at the same
point on the sheet ${\mathcal D}_H$ and leave it through the same
edges of different windows on ${\mathcal D}_H$. By construction of
$p^{(1)}(t)$ and $p^{(2)}(t)$, the endpoints of cycles  $\gamma_1$
and $\gamma_2$ have the same $\tau$-coordinate, as well as the
same $\eta,\mu$-coordinates (corresponding to $P_1\in \Gamma$).
Hence, the endpoints on ${\mathcal R}$ coincide, and by continuity
the corresponding ${\mathcal D}$-sheets can be identified.  \qed

\begin{figure}[h,t]
\begin{center}
\includegraphics[width=.6\textwidth]{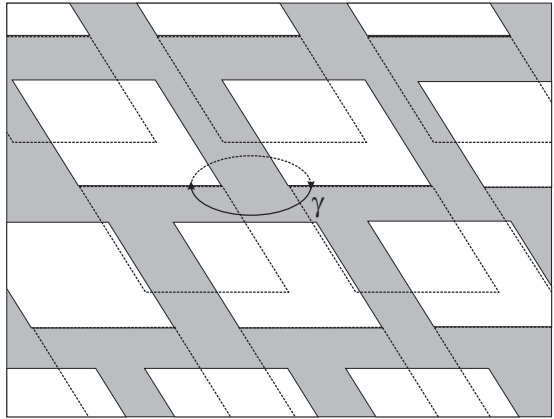}
\caption{ Window lattice on $\mathcal D_{H}$ (bold) and on
$\mathcal D_{HN}$ (dotted). The path leaves $\mathcal D_{H}$
through the $S$-side of a window (bold arc) and enters $\mathcal
D_{HN}$ , then it continues on $\mathcal D_{HN}$ (dotted arc) and
leaves the sheet through the $N$-edge of a window, coming back to
$\mathcal D_{H}$} \label{passages.fig}
\end{center}
\end{figure}

As a corollary of the above proposition, we note that each of the
four $\mathcal D$-sheets adjacent to ${\mathcal D}_H$ is in turn
connected in the same way to other four sheets, one of them being
 the initial sheet ${\mathcal D}_H$. In particular,
when $\tau(P)$ leaves ${\mathcal D}_H$ through the $S$-edge of a
window $W_{ij}$, enters the sheet ${\mathcal D}_{H,N}$ and then
leaves it through the $N$-edge of another window $W'_{ij}$ in
${\mathcal D}_{H,N}$, then it returns to ${\mathcal D}_H$ (see the
illustration in Figure \ref{passages.fig}). It is then natural to
label the ${\mathcal D}$-sheets connected to ${\mathcal D}_{HN}$
as ${\mathcal D}_{HNN}$, ${\mathcal D}_{HNE}$, ${\mathcal
D}_{HNW}$, and ${\mathcal D}_{HNS}={\mathcal D}_H$.

\vskip .2cm \noindent Finally, we also state the important

\begin{lem} The passages between $\mathcal D$-sheets in the ``horizontal'' ($E$, $W$)-direction and
the ``vertical'' ($N$, $S$)-direction commute. That is, the
following sheets coincide:
$$
{\mathcal D}_{HNE}={\mathcal D}_{HEN} , \quad {\mathcal
D}_{HNW}={\mathcal D}_{HWN}, \quad {\mathcal D}_{HSE}={\mathcal
D}_{HES} , \quad {\mathcal D}_{HSW}={\mathcal D}_{HWS} .
$$
\end{lem}

\vskip .2cm
 \noindent\textbf{Proof of Lemma.}
\vskip .2cm
  Indeed, the windows on
${\mathcal D}_{HNE}$ and on ${\mathcal D}_{HEN}$ are obtained from
those on ${\mathcal D}_H$ by shift by the same vector $V_1-V_3$,
hence ${\mathcal D}_{HNE}$ and ${\mathcal D}_{HEN}$ can be
identified. Similar arguments apply to identify the other pairs of
sheets. \qed

As a result, the Riemann surface $\mathcal R$ can be described as
a ${\mathbb Z}^2$-countable set of $\mathcal D$-sheets, that is
$${\mathcal R}= \{\cup \, {\mathcal D}_{ij}\mid i,j\in {\mathbb
Z}\}$$ where $i,j$ enumerate the passages in the ``horizontal''
and ``vertical'' directions respectively. Hence, travelling on
$\mathcal R$ can be encoded as a path (sequence) on a ${\mathbb
Z}^2$-lattice $\Lambda$.


\subsection{Branching of the projection $\mathcal R\mapsto\mathbb C=\{\tau\}$.}
Note that each corner of a window on a $\mathcal D$-sheet is
connected to several different sheets, hence the corners are
branch points of the projection $\Pi\, : \,{\mathcal R}\mapsto
{\mathbb C}=\{\tau\}$. In particular, consider the corner $\tau_*$
between the $E$- and $S$-edges of $W_{ij}$ on ${\mathcal D}_{H}$.
Let $\gamma_*$ be a path on $\mathcal R$ that starts on ${\mathcal
D}_{H}$ and whose projection to $\{\tau\}$ is a sufficiently small
circle around $\tau_*$ with a counter-clockwise orientation. Then
$\gamma_*$ leaves ${\mathcal D}_{H}$ through the $S$-side of a
window and visits the  sequence of sheets
$$
{\mathcal D}_{H} \to {\mathcal D}_{HN} \to {\mathcal D}_{HNE}
(={\mathcal D}_{HEN}) \to {\mathcal D}_{HE} \to {\mathcal D}_{H}.
$$
Thus $\tau_*$ connects four $\mathcal D$-sheets and four edges of
windows on them, and the path $\gamma_*$ closes after making 3
complete rotations about $\tau_*$. This gives a geometric
interpretation of the cubic root branching of functions
$\eta(\tau)$ and $\mu(\tau)$ in (\ref{fraction}). The other
 window corners have the same properties.
Note that there is no contradiction between the fact that each
branchpoint $\tau_*$ is of the {\em third order} but connects {\em
four} $\mathcal D$-sheets together. The reason is that the
$\mathcal D$-sheets are not copies of the whole complex plane
$\mathbb C$, and due to the presence of the windows it is not
possible to turn by $2 \pi$ around $\tau_*$ on a $\mathcal
D$-sheet.

\begin{figure}
\begin{center}
\includegraphics[width=.7\textwidth]{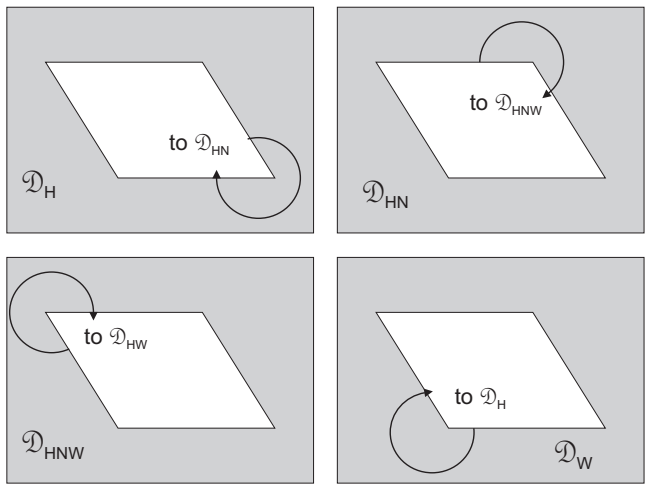}
\caption{A closed path on $\mathcal R$ around cubic branch point.
} \label{3-branching}
\end{center}
\end{figure}
Due to the incommensurability of the size of the windows (the
modulus of the forbidden periods $|V_1|$ and $|V_3$) with the
shift vectors that generate the windows lattice (the allowed
periods $V_2$ and $V_4$), the projection of the corners of all the
windows in all the sheets of $\mathcal R$ (i.e., the branch points
of $\Pi$) is a dense set in $\mathbb C$. However, as we have seen
above, on each of the infinite ${\mathcal D}$-sheets of $\mathcal
R$ the branch points are isolated and they form a regular lattice.

\subsection{Duality.} The ${\mathbb Z}^2$-lattice $\Lambda$
introduced above has a much richer structure. More precisely, one
can extend it to a cell structure $\bar\Lambda$ on ${\mathbb
R}^2$: the adjacent vertices of the lattice are regarded as
vertices of square cells $\Delta_{kl}$ on ${\mathbb R}^2$.

There is a remarkable duality between the objects of ${\mathcal
R}$ and $\bar\Lambda$: a 2-dimensional $\mathcal D$-sheet in
$\mathcal R$ corresponds to a vertex of $\Lambda$, the edges of
the windows connecting two different $\mathcal D$-sheets of
${\mathcal R}$ correspond to the segments joining the two
corresponding vertices. Finally, a branchpoint $\tau^*$ of
${\mathcal R}$ which connects four $\mathcal D$-sheets corresponds
to the square cell $\Delta_{kl}$ in $\bar\Lambda$ defined by the
four vertices which correspond to the sheets.

\subsection{The surface $\mathcal R$ as the universal covering of
the theta-divisor.} Now, instead of the single hyperelliptic
integral (\ref{int_main}), consider the vector of holomorphic
integrals
$$
 (u_1, u_2)^T= \int_{\infty}^P \left( \frac {d\eta} \mu, \frac {\eta\, d\eta} \mu \right)^T,
\qquad P\in \Gamma,
$$
which has 4 independent vectors of periods in $ {\mathbb C}^2
(u_1, u_2)$ with respect to the cycles $a_1,\dots,b_2$. As it is
known from the theory of Abelian varieties, the map ${\mathcal
A}\, :\, P \mapsto (u_1, u_2)^T$ describes a well defined smooth
embedding of $\Gamma$ into the Jacobian variety Jac$(\Gamma)=
{\mathbb C}^2 /\Lambda$, where $\Lambda$ is the lattice generated
by the above period vectors. The image of the embedding is known
to be a one-dimensional analytic subvariety $\Theta' \subset$
Jac$(\Gamma)$, a translate of the theta-divisor defined as the
zero locus of the Riemann theta-function $\theta(u_1, u_2)$ of the
curve $\Gamma$.

Let $\widetilde\Theta \subset {\mathbb C}^2$ be the universal covering of $\Theta$.
Note that it can also be regarded as an infinitely sheeted
ramified covering of ${\mathbb C}=\{u_1 \}$ and that the first
coordinate $u_1$ jointly with the point $P\in \Gamma$ define a
point of $\widetilde\Theta$ uniquely. It follows that by
identifying $u_1$ with the complex coordinate $\tau$ in
(\ref{int_main}), one can also identify the universal covering
$\widetilde\Theta$ with our Riemann surface $\mathcal R$: two
points on both complex surfaces are identified if and only if they
are projected to the same $u_1=\tau$ and correspond to the same
point on $\Gamma$.

\subsection{Periods of the pentagonal curve.}

The existence of a group of automorphisms of the curve $\Gamma_5=
\{ \mu^2=E-\eta^5\}$ implies some special relations between its
periods, which we did not find in the literature and indicate
here.

\begin{prop} \label{periods-5}
For the chosen basis of cycles on $\Gamma_5$, the local dependence
of the periods on the complex energy is as follows
\begin{equation} \label{trivial}
 V_i (E)= E^{-3/10} V_i^0  ,\qquad i=1,\dots,4,
\end{equation}
where $V_i^0$ are the periods for $E=1$ given by
\begin{equation}  \label{5-periods}
\begin{aligned}
V_{1}^0 &= 2\int_\infty^{1} \frac{d\eta}{\sqrt{1-\eta^5}}\,
\cong\,
1.819915 \,{\rm i}\\
V_{2}^0 & = -\frac {3+\sqrt{5}}{2}\, V_{1}^0, \\
V_{3}^0 &= - {\rm e}^{-{\rm i}\pi /5} V_{1}^0, \\
V_{4}^0 & =-\frac {3+\sqrt{5}}{2}\, V_{3}^0,
\end{aligned}
\end{equation}
\end{prop}

Note that when $E$ varies along a loop embracing the origin and
returns to its original value, the curve $\Gamma_5$ undergoes the
automorphism $\rho$ that rotates its finite branch points by angle
$2\pi/5$, whereas the periods undergo a monodromy: they are all
multiplied by $\exp(-3\pi i/5)$ and, on the other hand, become
linear combinations of $V_{1}^0, \cdots, V_{4}^0$ with integer
coefficients. Proposition (\ref{5-periods}) implies that for any
complex energy $E$ the following relations hold
\begin{equation} \label{coll}
\frac{V_2}{V_1}=\frac{V_4}{V_3}=-\frac {3+\sqrt{5}}{2}\, , \quad
|V_1|=|V_3|, \quad |V_2|=|V_4|.
\end{equation}
The two forbidden (allowed) periods have the same length, moreover
the forbidden period $V_1$ ($V_3$) on $\mathbb C$ is proportional
to the allowed period $V_2$ ($V_4$). Since the proportionality
constant is an irrational number, the period vectors are not
commensurable, as was also claimed in Remark 2.1.

It is worth noting that the pentagonal curve $\Gamma_5$ is rather
special, and the form of the windows on ${\mathcal D}$ for a
generic genus 2 curve $\{\mu^2=P_5(\eta)-E\}$ depends on $E$ in a
nontrivial way: $V_i$ form a fundamental basis of solutions of a
fourth order Picard--Fuchs ODE with the independent variable $E$.
\vskip 0.3cm \noindent\textbf{A sketch of the proof of Proposition
\ref{periods-5}} \vskip 0.3cm
 Under the subsitution $\eta=E^{1/5} \lambda$, the holomorphic differential
$d\eta/\mu$ takes the form $E^{-3/10}\,
d\lambda/\sqrt{1-\lambda^5}$, while on the $\lambda$-plane the
branch points and the integration cycles are the same as on
$\eta$-plane for $E=1$. This implies the relations
(\ref{trivial}). The calculation of the periods $V_{1}^0, \cdots,
V_{4}^0$  follow
 the ideas of \cite{Bolza}. More specifically, the monodromy
transformation described above implies that the periods $V_i^0$
must satisfy a system of homogeneous linear equations with integer
coefficients
$$
\exp(-3\pi i/5) V_{i}^0 = \alpha_{i1} V_1 +\cdots + \alpha_{i4}
V_4, \qquad \alpha_{ik} \in {\mathbb Z} .
$$
Expressing the deformed cycles $\rho(a_1), \dots, \rho(b_2)$ in
terms of the original cycles on $\Gamma_5$ for $E=1$, it is
possible to calculate the coefficients $\alpha_{ik}$. Finally,
solving this system and numerical integration of $V_1^0$ yields
expressions (\ref{5-periods}). \qed

 \section{Circular Travelling on the Surface $\mathcal R$}\label{sec:circular}
 As mentioned in Section 1, one can
consider a complex solution of the system (1.12) with $t\in
{\mathbb R}$ as a smooth path $\gamma (t)$ on the infinitely
sheeted surface $\mathcal R$ corresponding to the curve $\Gamma$
such that the projection $\Pi (\gamma(t))$ is the circle $\mathcal
C$ on the $\tau$-plane parameterized by (\ref{eq:trick1}) and thus
having radius $1/\omega$.

\subsection{A geometrical algorithm}

Comparing the quadrature \eqref{int5}  and the integral
(\ref{int_main}), we find that for initial complex conditions
$\zeta_0, \zeta'_0$, the $\tau$-coordinate of the starting point
$\gamma_0=\gamma (0)$ is
\begin{equation}\label{tau0}
 \tau_0 = \int_\infty^{\zeta_0} \frac{d\eta}{\sqrt{2}\sqrt{E-\eta^5}} ,
\end{equation}
where the energy $E$ is defined in \eqref{eq:E} and the sign of
the root is chosen in accordance with the sign of $\zeta'(0)$.

Geometrically, $\gamma_0$ can be chosen on any ${\mathcal
D}$-sheet by adding $\tau_0$ to a corner of any window\footnote{If
for a chosen corner $\tau_0$ appears inside  a window, then
$\tau_0$ should be added to another corner.}. By convention, the
chosen initial sheet is labelled as ${\mathcal D}_{00}$.

Let ${\mathcal C}_0$ be the preimage of the circle ${\mathcal C}$
on ${\mathcal D}_{00}$. Naturally, if ${\mathcal C}_0$ does not
cross any window on ${\mathcal D}_{00}$,
 then the path  $\Pi^{-1}\, {\mathcal C}$ on $\mathcal R$
remains on ${\mathcal D}_{00}$ and the solution $\zeta(t)$ of
(\ref{eq:zeta(t)}) is periodic with the fundamental period $T=2\pi
/\omega$. In other cases the path $\gamma (t)$ gives rise to a
finite or infinite sequence on the ${\mathbb Z}^2$-lattice $\{\cup
\, {\mathcal D}_{ij}\mid i,j\in {\mathbb Z}\}$
according to the following algorithm:

\begin{description}

\item{1)} Given position of the circle ${\mathcal C}_0$ on
${\mathcal D}_{00}$ with the initial point $\gamma_0\in {\mathcal
C}_0$, one determines the points of its intersection with the
edges of ${\mathcal W}_{ij}$. Among them one chooses the point
$\gamma_1$, which is closest to $\gamma_0$ in the counterclockwise
direction on ${\mathcal C}_0$.

\item{2)} Given $\gamma_1$ and the edge containing it, one finds
the point $\gamma_1'$ on the opposite edge of the same window such
that
$$
\gamma_1'=\gamma_1 \pm V_1 \quad \mbox{or } \quad
\gamma_1'=\gamma_1\pm V_3.
$$
Then $\gamma (t)$ passes to the new sheet $ {\mathcal D}'
={\mathcal D}_{0,\pm 1}$ (respectively, ${\mathcal D}' ={\mathcal
D}_{\pm 1,0}$) with $\gamma_1'$ as the starting point on
${\mathcal D}'$.

\item {3)} On the new sheet ${\mathcal D}'$ one constructs the
circle  ${\mathcal C}_1$ by translating ${\mathcal C}_0$ by $\pm
V_1$ (respectively $\pm V_3$) so that  $\gamma_1'\in {\mathcal
C}_1$. One then determines the intersections of ${\mathcal C}_1$
with edges of windows on ${\mathcal D}'$ and chooses the
intersection point $\gamma_2$, which is closest to $\gamma_1'$ in
the clockwise direction on ${\mathcal C}_1$.

Then the procedure repeats as in step 2) with ${\mathcal D}_{00}$
and $\gamma_1$ replaced by ${\mathcal D}'$ and $\gamma_2$.

\item{4)} If after $n$ iterations one returns to ${\mathcal
D}_{00}$, then the corresponding circle ${\mathcal C}_{n-1}$
necessarily coincides with ${\mathcal C}_0$. If the segment of
${\mathcal C}_0$ between $\gamma_n$ and $\gamma_{n+1}$ contains
$\gamma_0$, then the procedure stops and $\gamma(t)$ is closed.
Otherwise, the algorithm continues on step 1) with $\gamma_0$
replaced by $\gamma_n$.
\end{description}

This geometric algorithm for travelling on the fundamental domain
includes the essential properties of the flow defined by the
system (\ref{eq:xyk4}) and allows to predict the period of the
orbit without  numerical integration.


\subsection{Motion on a fundamental domain $\tilde {\mathcal
D}$.} 
\begin{figure}[h]
\begin{center}
\begin{tabular}[h]{cc}
\begin{tabular}{c}
\psfig{figure=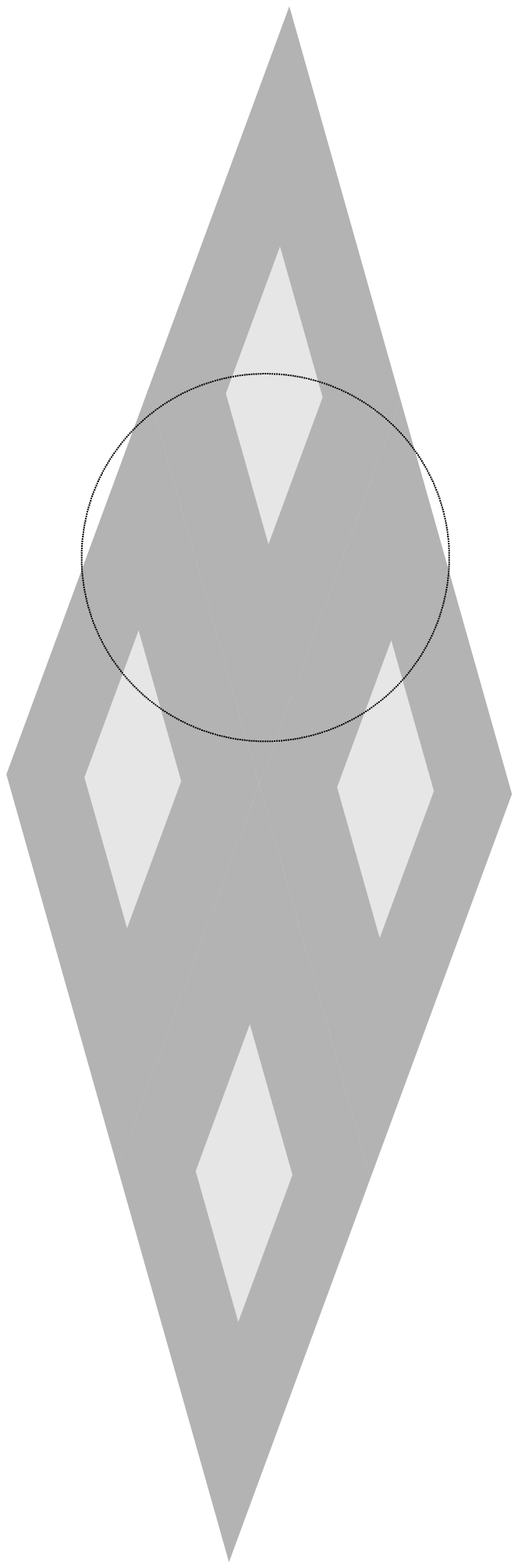,height=2.2in}\\[-8pt] {\scriptsize{\bf (a) } }
\end{tabular} &
\begin{tabular}{c}
\psfig{figure=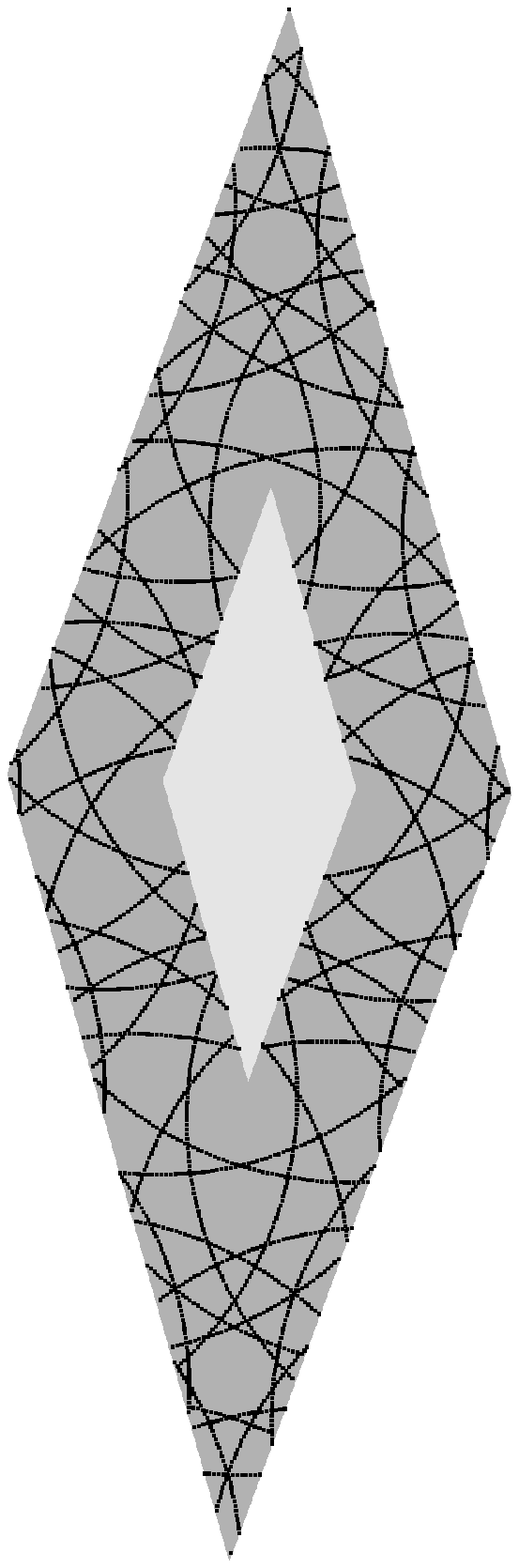,height=2.2in}\\[-8pt] {\scriptsize{\bf (b) } }
\end{tabular}\\
\begin{tabular}{c}
\psfig{figure=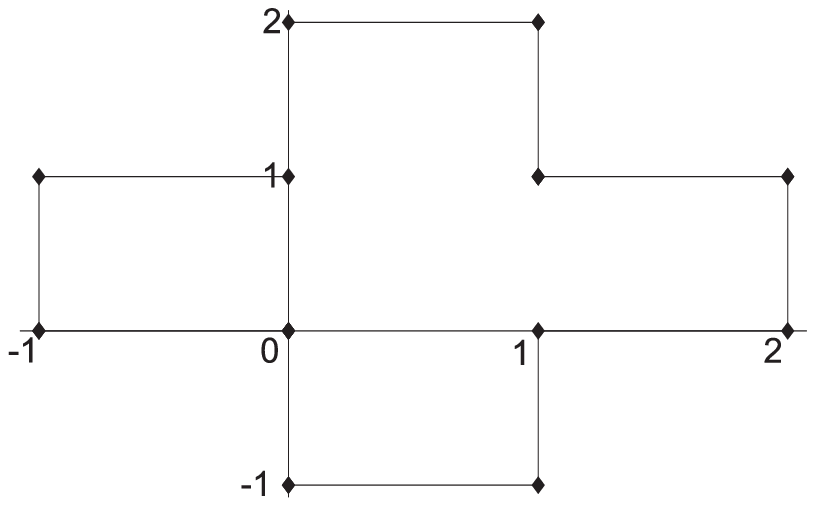,width=1.9in}\\[-8pt] {\scriptsize{\bf (c) } }
\end{tabular} &
\begin{tabular}{c}
\psfig{figure=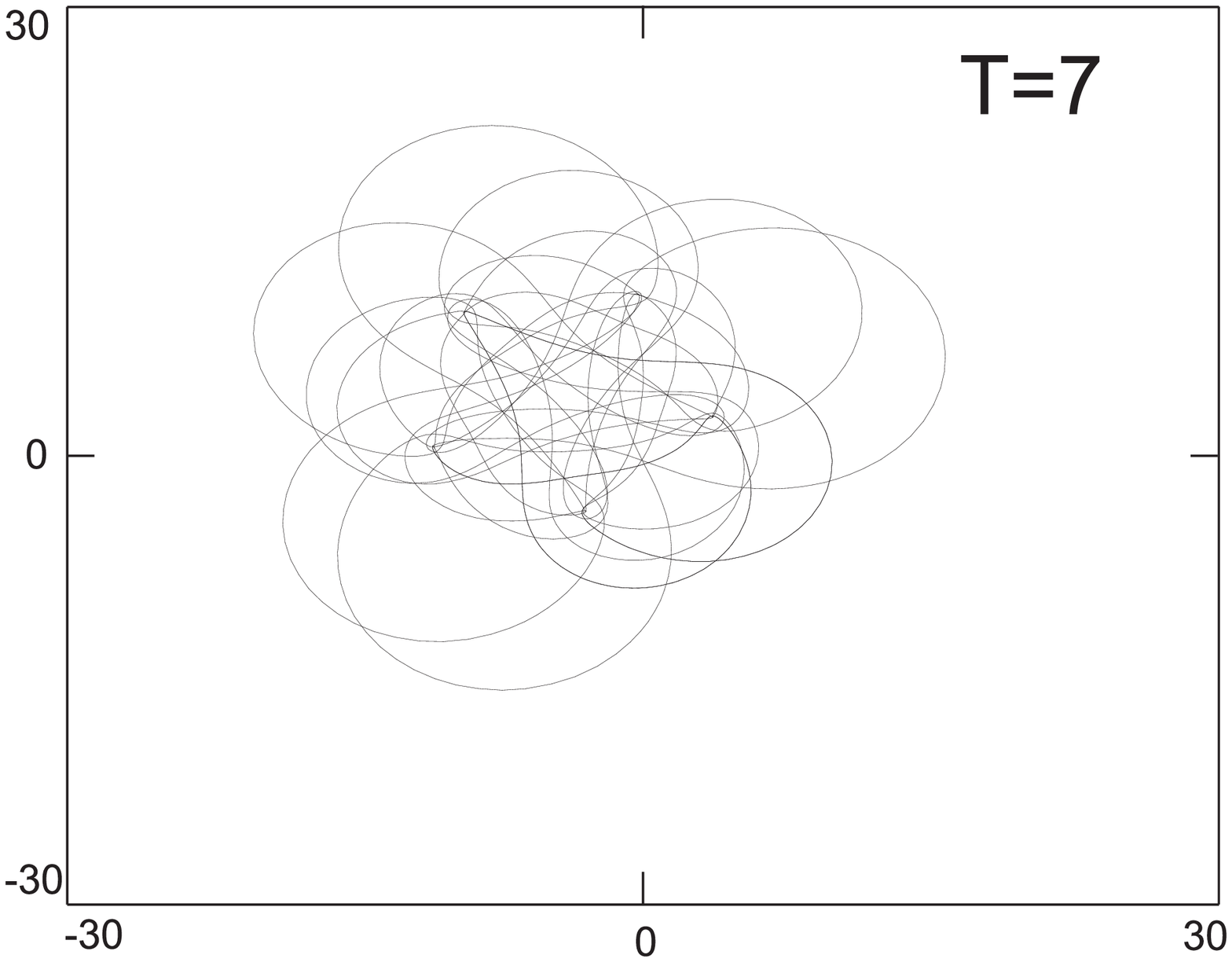,width=1.9in}\\[-8pt] {\scriptsize{\bf (d) } }
\end{tabular}
\end{tabular}
\end{center}
\caption{{\bf (a)} Circle $\mathcal C_0$  on the initial sheet
$\mathcal D_{00}$. {\bf (b)} Path $\tilde \gamma(t)$ on the
fundamental domain $\tilde D$. {\bf (c)} Sequence of sheets of
$\mathcal R$ visited by $\gamma(t)$. {\bf (d)} Orbit of $\zeta(t)$
obtained by numerical integration of \eqref{eq:zeta(t)} with
period $7\,T$.}\label{fig:example}
\end{figure}

There is an equivalent description of the procedure. Namely,
let $\hat {\mathcal D}$ be a fundamental parallelogram of allowed
periods $V_2$ and $V_4$ on $\mathcal D$ that contains a window
${\mathcal W}^*$ completely inside. Then $\hat{\mathcal D}$ does
not have any other windows inside and there is a natural bijection
between the fundamental domain $\tilde {\mathcal D}= \hat{\mathcal
D}\setminus {\mathcal W}^* \subset {\mathcal D}$ and the curve
$\Gamma$. A smooth path $\gamma(t)= \Pi^{-1}\, {\mathcal C}$ on
$\mathcal R$ induces a smooth path $P(t)$ on $\Gamma$ and a
generally discontinuous path $\tilde\gamma(t)$ on $\tilde
{\mathcal D}$ that consists of segments of translates of
${\mathcal C}_0$: when $\tilde\gamma(t)$ crosses an exterior edge
of $\tilde {\mathcal D}$ at a point $\gamma^*$, it reappears on
the point $\gamma^{**}$ on the opposite exterior edge of the
domain, since both points correspond to the same point on
$\Gamma$. Similarly, on crossing an edge of the interior window of
$\tilde {\mathcal D}$, the trajectory continues from the opposite
edge of the window as described in step 2) of the  algorithm.
\vskip 0.3cm \noindent\textbf{Example} \vskip 0.3cm
 For the pentagonal curve with the complex energy
$E=1110+i\, 1452 $, a fragment of the initial sheet ${\mathcal
D}_{00}$ and the corresponding fundamental domain $\tilde
{\mathcal D}$ are presented in Figure \ref{fig:example}a), and
\ref{fig:example}b) respectively. Here we also show an example of
the circle ${\mathcal C}_0$ on ${\mathcal D}_{00}$, which
generates a periodic path $\tilde\gamma(t)$ in $\tilde {\mathcal
D}$ with period $7T$. The corresponding periodic sequence on the
${\mathbb Z}^2$-lattice is
\begin{align*}
\overline{(0, 0)} & \to  (-1, 0) \to  (0, 0) \to  \overline{(-1, 0)} \to  (-1, 1) \to  \overline{(0, 1)}  \to (0, 2) \to \\
 \overline {(1, 2)} & \to  (1, 1) \to  ( 1, 2) \to  (1, 1) \to  \overline {(2, 1)} \to  (1, 1) \to (2, 1)  \to \\
 (2,0) & \to  \overline {(1,0)} \to (1,-1) \to \overline{(0,-1)} \to (0,0) \to (0,-1) \to \overline {(0,0)} .
\end{align*}
where the bars mark the ${\mathcal D}$-sheets visited when the
projection $\Pi \gamma(t)$ performs a complete turn. Applying the
above algorithm to different initial conditions, we obtained a
great variety of periodic trajectories. In all cases we observed a
perfect coincidence of the motion in $\tilde {\mathcal D}$ with
the results obtained by the numeric integration of the equations
(\ref{eq:zeta(t)}) and (\ref{eq:xyk4}).

\subsection{Aperiodic and isochronous trajectories}
The algorithm described above does not cover the cases when, at a
certain step, $\gamma(t)$ arrives precisely at the corner of a
window. Since the corners are  branch points of $\mathcal R$
connecting 4 different ${\mathcal D}$-sheets, it is not possible
to choose uniquely the next sheet, and at this step the procedure
should terminate. Note that in this case the corresponding complex
solution $z(t)$ escapes to infinity in finite time.

We also observe that if an aperiodic path $\gamma(t)$ exists, then
it must pass arbitrarily close to some branch point on $\mathcal
R$ for sufficiently large time. The following obvious properties
hold.

\begin{prop} \label{non-period}
 If the path $\gamma(t)=\Pi^{-1}\, {\mathcal C}$ is not
periodic, then  it visits an infinite set of ${\mathcal
D}$-sheets, and for any $\delta>0$ there exist a time $t^*$ and a
branch point $\tau^*\in {\mathcal R}$ such that
$|\gamma(t^*)-\tau^*|<\delta$.
\end{prop}

In other words, if an aperiodic orbit exists, then it has
sensitive dependence on initial conditions: if $\gamma_0$ yields a
non-periodic path $\gamma(t)$ on $\mathcal R$ and its arbitrarily
small variation $\tilde \gamma_0$ results in a path
$\tilde\gamma(t)$, then there always exist a finite $t^*$ such
that for $t<t^*$ both paths are very close to each other, but for
$t>t^*$ they pass by a branch point $\tau^*$ on different sides.
As a consequence $\gamma(t)$ and $\tilde \gamma(t)$ enter
different ${\mathcal D}$-sheets and their future behavior is
completely different. \vskip .2cm \noindent\textbf{Proof of
Proposition \ref{non-period}}\vskip .2cm

Suppose that there exists a non-periodic path $\gamma(t)$ visiting
a finite number of ${\mathcal D}$-sheets, then some of these
sheets are passed an infinite number of times. On each of them the
corresponding translate of $\mathcal C$ is the same and consists
of a finite number of arcs. Hence at least one of the arcs is
passed more than once and therefore $\gamma(t)$ must be periodic.
For the second part, since $\gamma(t)$ visits an infinite number
of ${\mathcal D}$-sheets, its image in the fundamental domain
$\tilde {\mathcal D}$ consists of an infinite set of segments of
translates of $\mathcal C$ by non-commensurable periods
$V_1,\dots, V_4$. Hence the arcs cover a dense set on
$\tilde{\mathcal D}$, and in particular they pass arbitrarily
close to any corner of the window in $\tilde{\mathcal D}$. \qed

\vskip .2cm

On the contrary, it is easy to prove that the periodic solutions
of (\ref{eq:zeta(t)}) are {\em isochronous}, i.e. , for any
initial data $\zeta(0),\zeta'(0)$ that give rise to a periodic
orbit, there exists an open set of initial data that contains
$\zeta(0),\zeta'(0)$ such that all trajectories starting on it are
also periodic, {\em with the same period}. This result follows
simply from the fact that branch points are isolated on $\mathcal
R$ and each sheet is connected to a finite number of sheets.

Although one cannot exclude that non-periodic trajectories of the
(\ref{eq:zeta(t)}) and (\ref{eq:xyk4}) may exist, we failed to
find any concrete scenario of an aperiodic motion on ${\mathbb
Z}^2$ except the case of critical values of $E$, for which the
curve $\Gamma$ becomes singular. We formulate thus the following
conjecture.

\begin{conj} For any non-critical energy $E$ the system \eqref{eq:xyk4} possesses
only periodic trajectories or singular ones (that escape to
infinity in finite time) and almost all of them are periodic.
However, for any $k\in {\mathbb N}$, initial data can be found
such that the trajectory has period larger than $kT$, where
$T=2\pi/\omega$ is the fundamental period.
\end{conj}

\section{Numerical Results}\label{sec:numerical}

In this Section we want to test the validity of our results by
comparing them with a numerical integration of the equations of
motion.
\subsection{Results for $k=4$}
We study trajectories of dynamical system (\ref{eq:xyk4})
corresponding to different initial data. For each set of initial
data $\{x(0),y(0),\dot x(0), \dot y(0)\}$ we compute the
``energy'' $E$ using \eqref{initialdata} and \eqref{eq:E}. The
value of $E$ determines uniquely the four period vectors, which
are defined generically through the contour integrals
\eqref{periodvecs}. For the special case of the pentagonal curve
 the four periods vectors are simply given by the formula
\eqref{5-periods}. As explained in Section \ref{sec:inversion} the
 allowed periods $V_2$ and $V_4$ define a quasi-elliptic period
lattice on the plane while the forbidden  periods $V_1$ and $V_3$
define the windows. In the special case of the pentagonal curve
\eqref{coll} applies, so that each sheet of the Riemann surface
$\mathcal R$ can be viewed as a fundamental cell with a window
such as those depicted in Figure \ref{fig:fundcell}.

In order to apply the algorithmic procedure described in Section
\ref{sec:circular} to lift the circular path to the Riemann
surface $\mathcal R$ it is only necessary to calculate the
position of the circle relative to the windows (the radius of the
circle is always fixed for every initial data, and equal to
$1/\omega$), that is to say, calculating $\tau_0$ in
\eqref{eq:trick1} by performing the integral \eqref{tau0}.

We have considered four different sets of initial data. In all of
them the initial velocities $\dot x(0) = \dot y(0) =0$, while the
initial positions $\{x(0),y(0)\}$ are displayed in Table
\ref{tab1}. For simplicity we have also set $\omega=2\pi$ so that
the fundamental period $T=1$.

\begin{figure}[h]\label{fig:circles}
\begin{center}
\psfig{figure=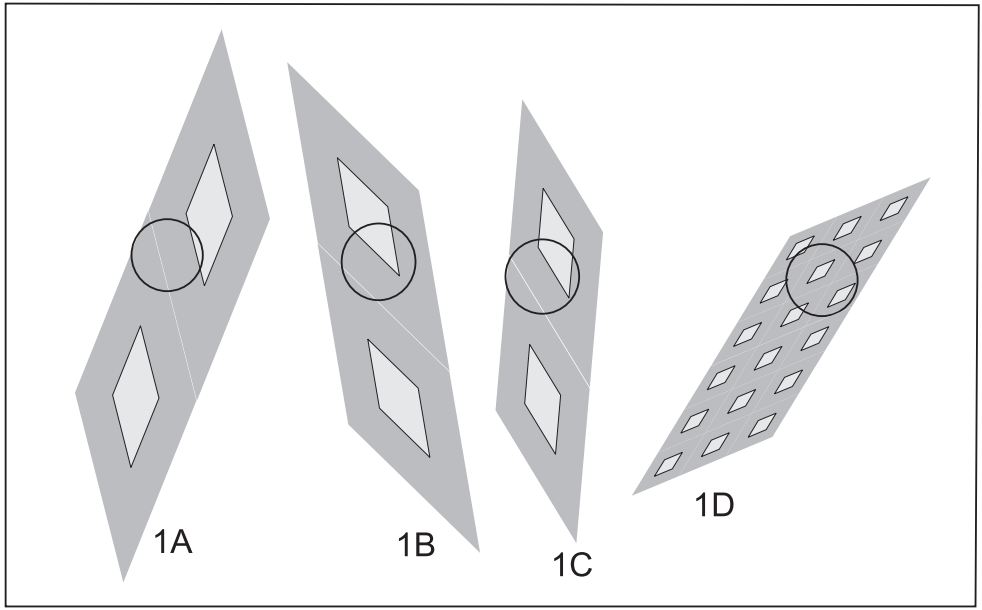,width=13cm}
 \caption{Evolution circle projected on the home sheet $\mathcal D_{00}$ for different initial data gathered in Table \ref{tab1} .}
\end{center}
\end{figure}

\begin{figure}[h]\label{fig:fundcell}
\begin{center}
\psfig{figure=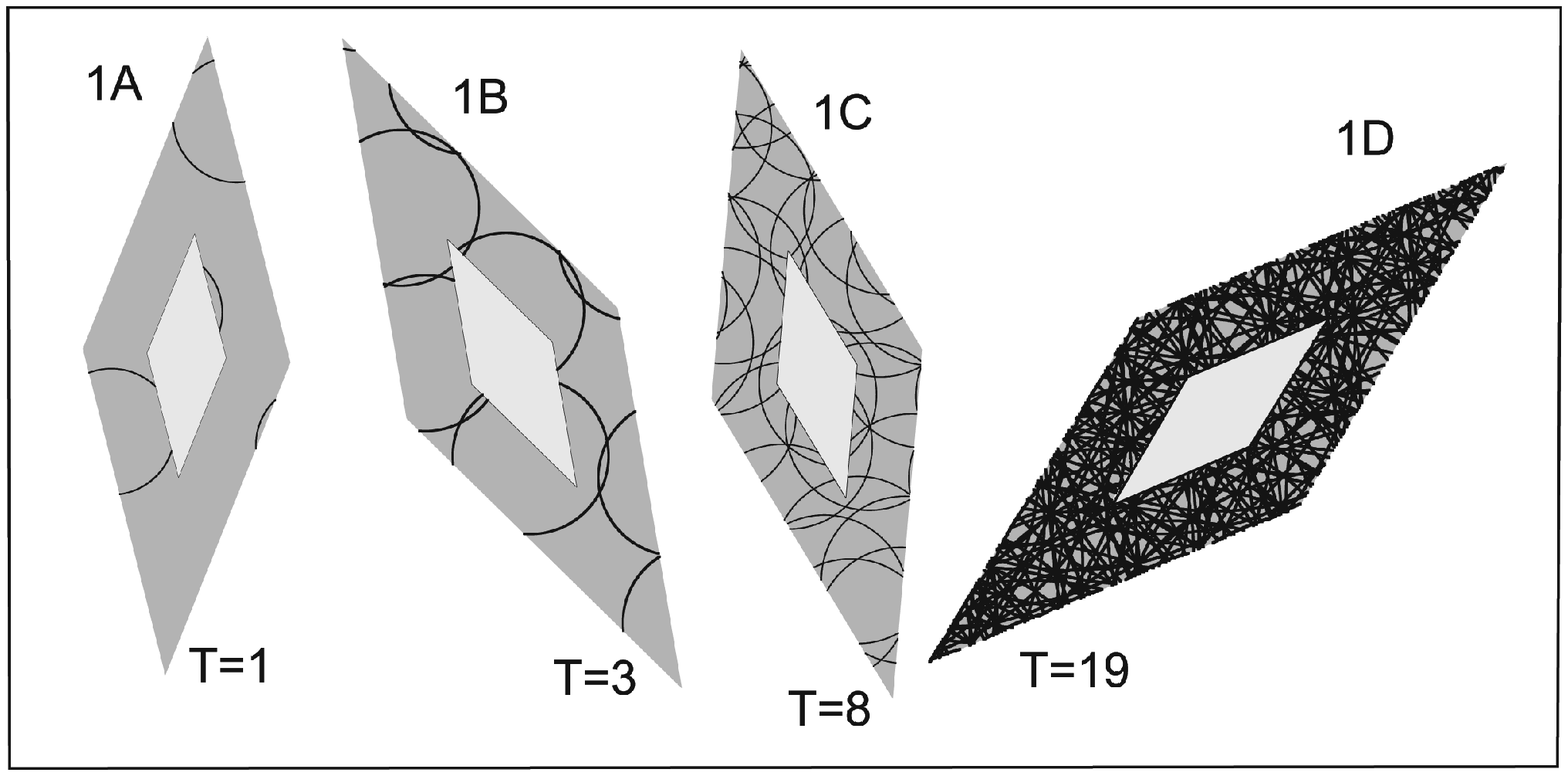,width=13cm}
 \caption{Path $\gamma(t)$ on the fundamental domain for different initial data gathered in Table \ref{tab1}.}
\end{center}
\end{figure}

\begin{table}[h]\caption{Different sets of initial data for $k=4$
\label{tab1}}\vskip 0.2cm
\begin{center}
\begin{tabular}{|c|c|c|c|c|}
  \hline
  Case & Initial data $(x_0,y_0)$ & $E$ & $\tau_0$ & Period \\
  \hline
 1A &$(1, 2)$ & $67.318 -73.092{\rm i}$ & $0.179+1.242{\rm i}$ & $1$ \\
 1B &$(2, 1.5)$  & $-112.73 -60.044{\rm i}$  & $-0.439+1.091{\rm i}$ & $3$ \\
 1C &$(1.8, 2.1)$  & $-52.998 - 215.31{\rm i}$ & $-0.141 + 0.988{\rm i}$ & $8$ \\
 1D &$(5.65, 2.1)$  & $-1888.3 +7576.3{\rm i}$ & $0.171 + 0.140{\rm i}$ & $19$ \\
  \hline
\end{tabular}
\end{center}
\end{table}
For each of the values above, the evolution circle on the window
lattice of the home sheet $\mathcal D_{00}$ has been displayed in
Figure \ref{fig:circles} . Note that the size of the windows
relative to the circle decreases as $|E|$ increases, see
\eqref{trivial}.

The algorithm described in Section \ref{sec:circular} to follow
the path on the Riemann surface $\mathcal R$  has been implemented
in Mathematica. For each of the cases 1A--1D the paths on
$\mathcal R$ are all periodic, and the periods predicted by the
algorithm are indicated in the last column of Table \ref{tab1}. In
Figure \ref{fig:fundcell} we show the discontinuous path
$\gamma(t)$ on the fundamental domain $\tilde {\mathcal D}$ for
each of these cases.

\begin{figure}[h]\label{fig:z2}
\begin{center}
\begin{tabular}[h]{cc}
\begin{tabular}{l}
\psfig{figure=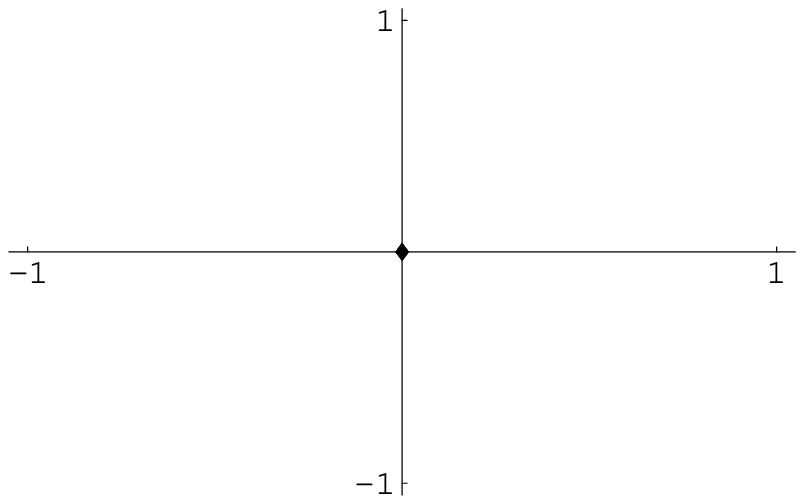,width=2.3in}
\end{tabular} &
\begin{tabular}{l}
\psfig{figure=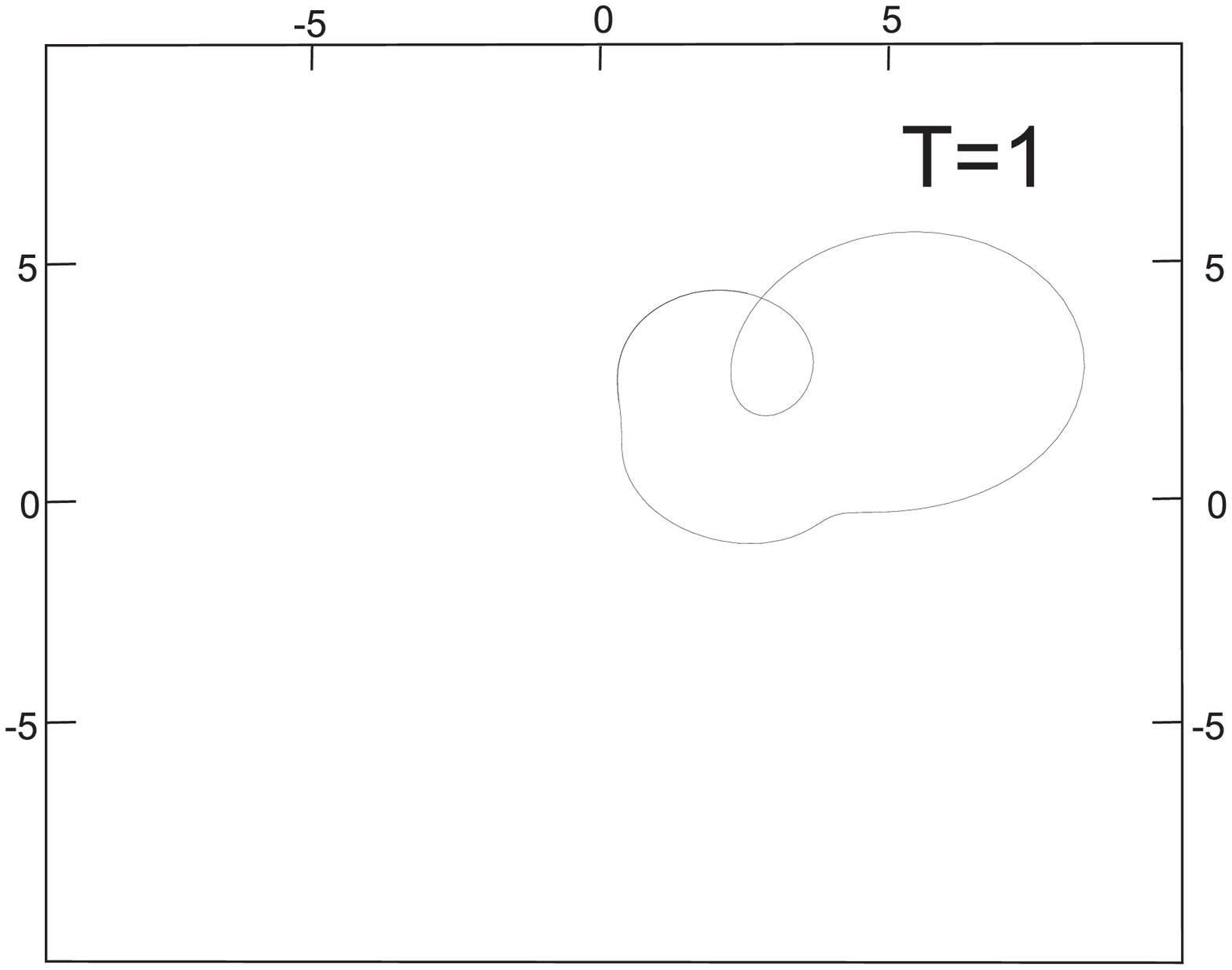,width=2.2in}
\end{tabular}
\\\\
\begin{tabular}{l}
\psfig{figure=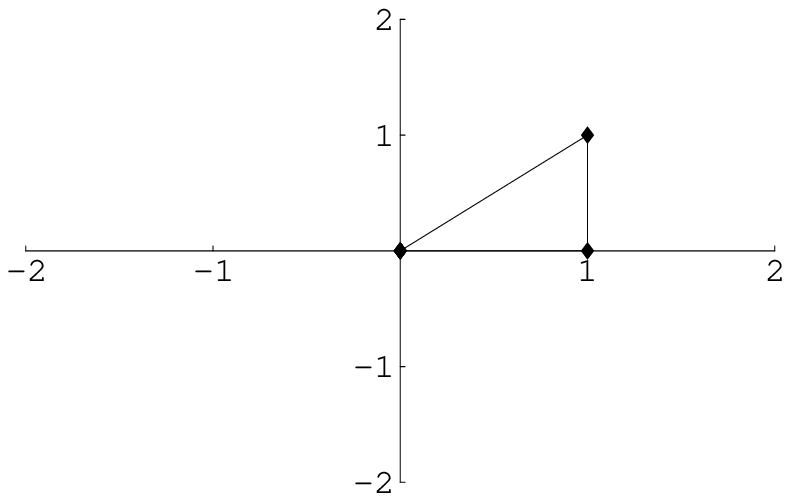,width=2.3in}
\end{tabular}
&
\begin{tabular}{l}
\psfig{figure=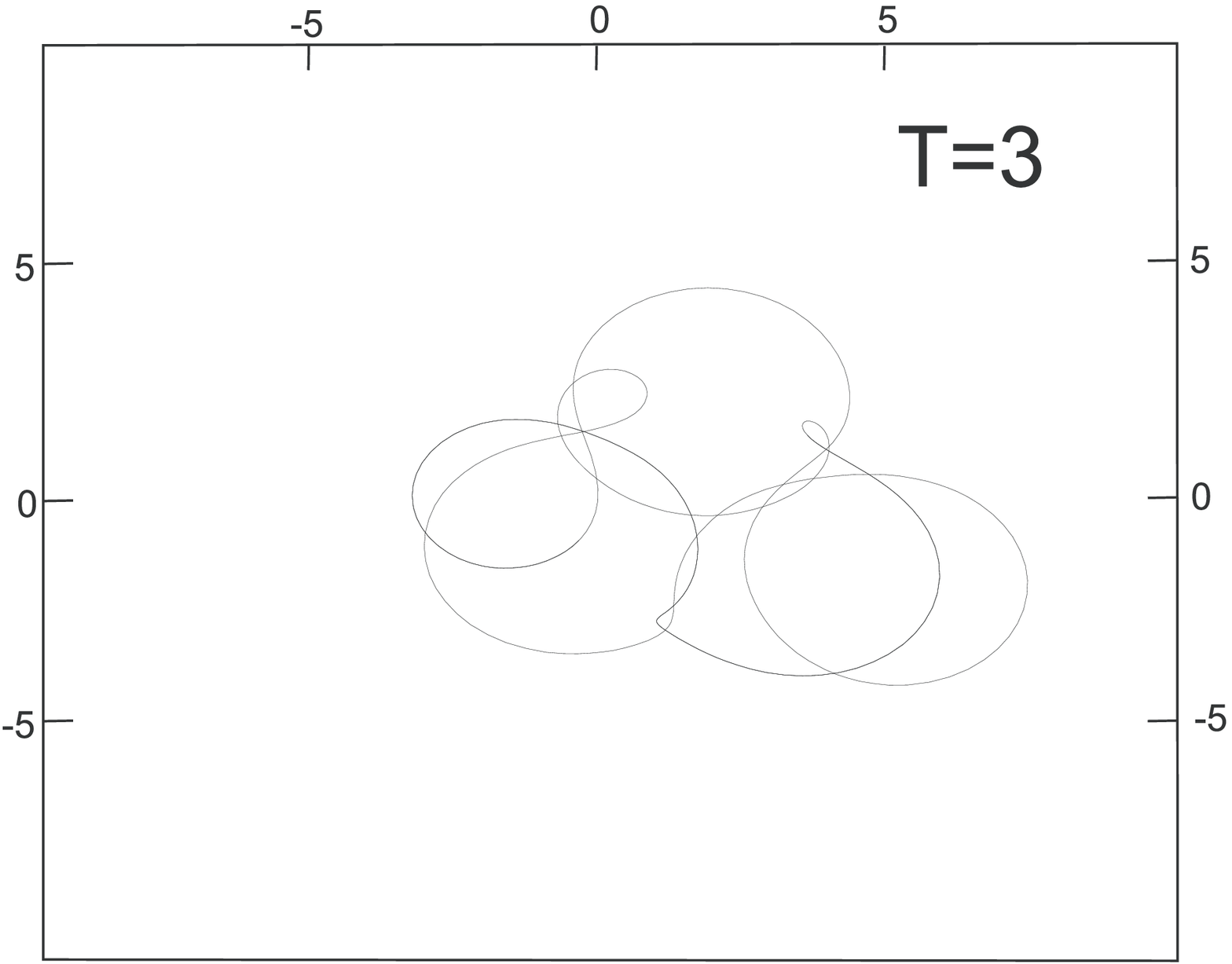,width=2.2in}
\end{tabular}
\\ \\ \begin{tabular}{l}
\psfig{figure=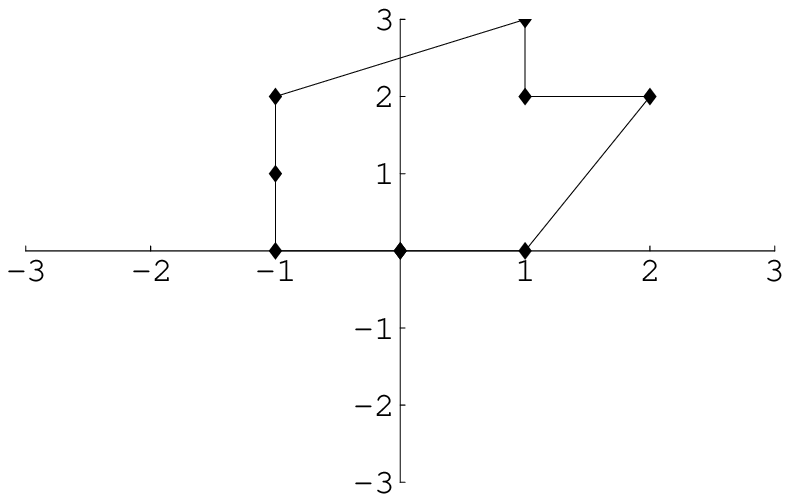,width=2.3in}
\end{tabular}
&
\begin{tabular}{l}
\psfig{figure=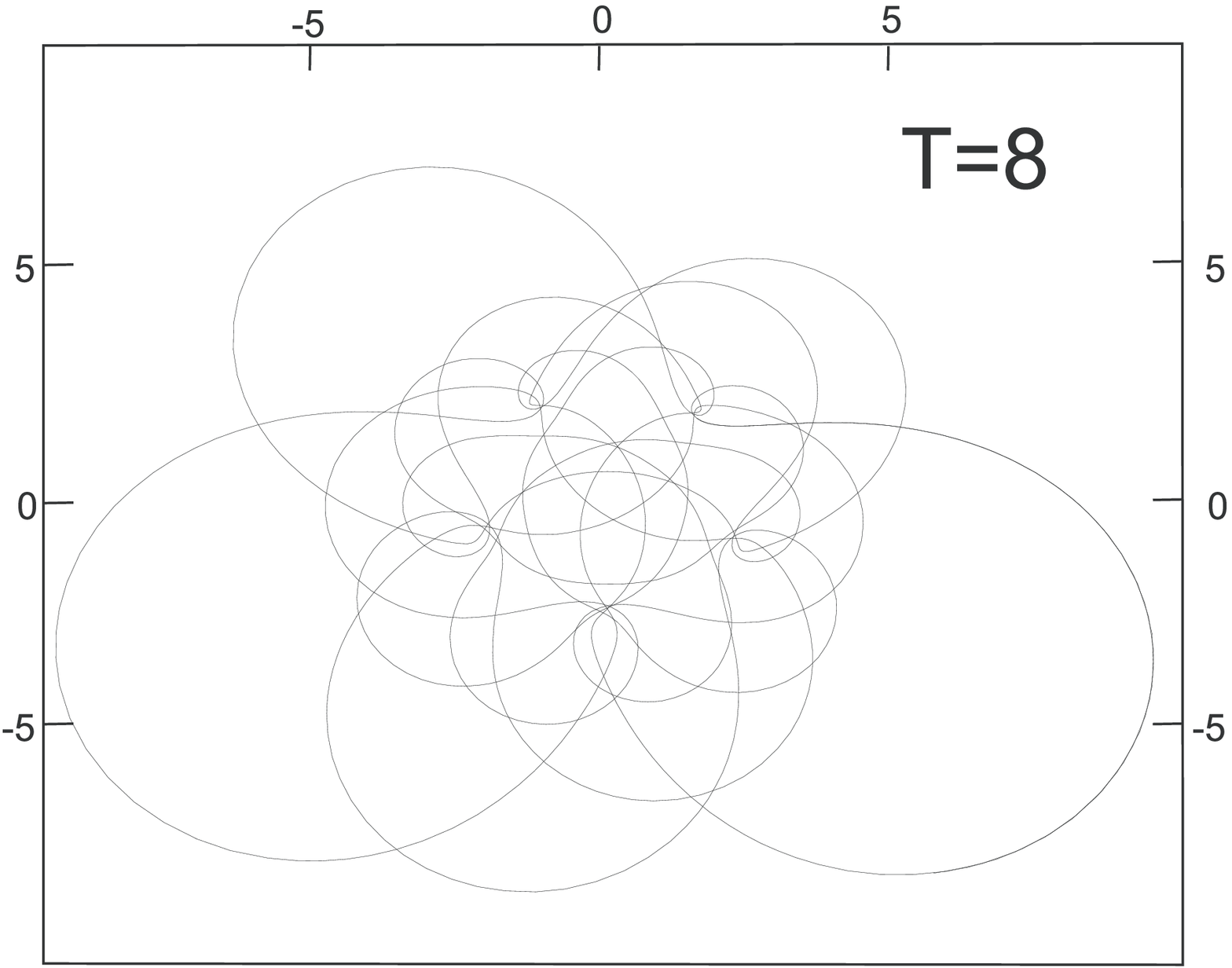,width=2.2in}
\end{tabular}
\\\\
\begin{tabular}{l}
\psfig{figure=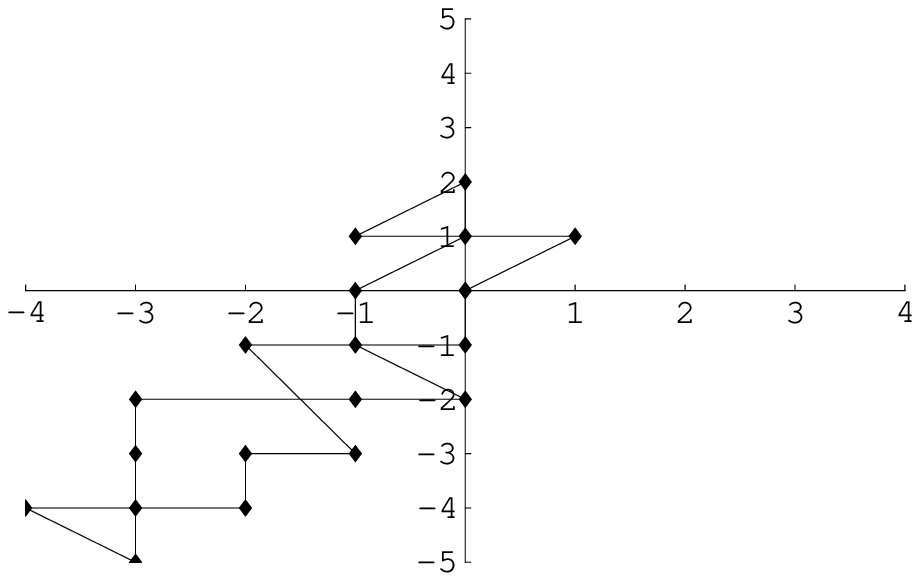,width=2.3in}
\end{tabular}
& \begin{tabular}{l}
\psfig{figure=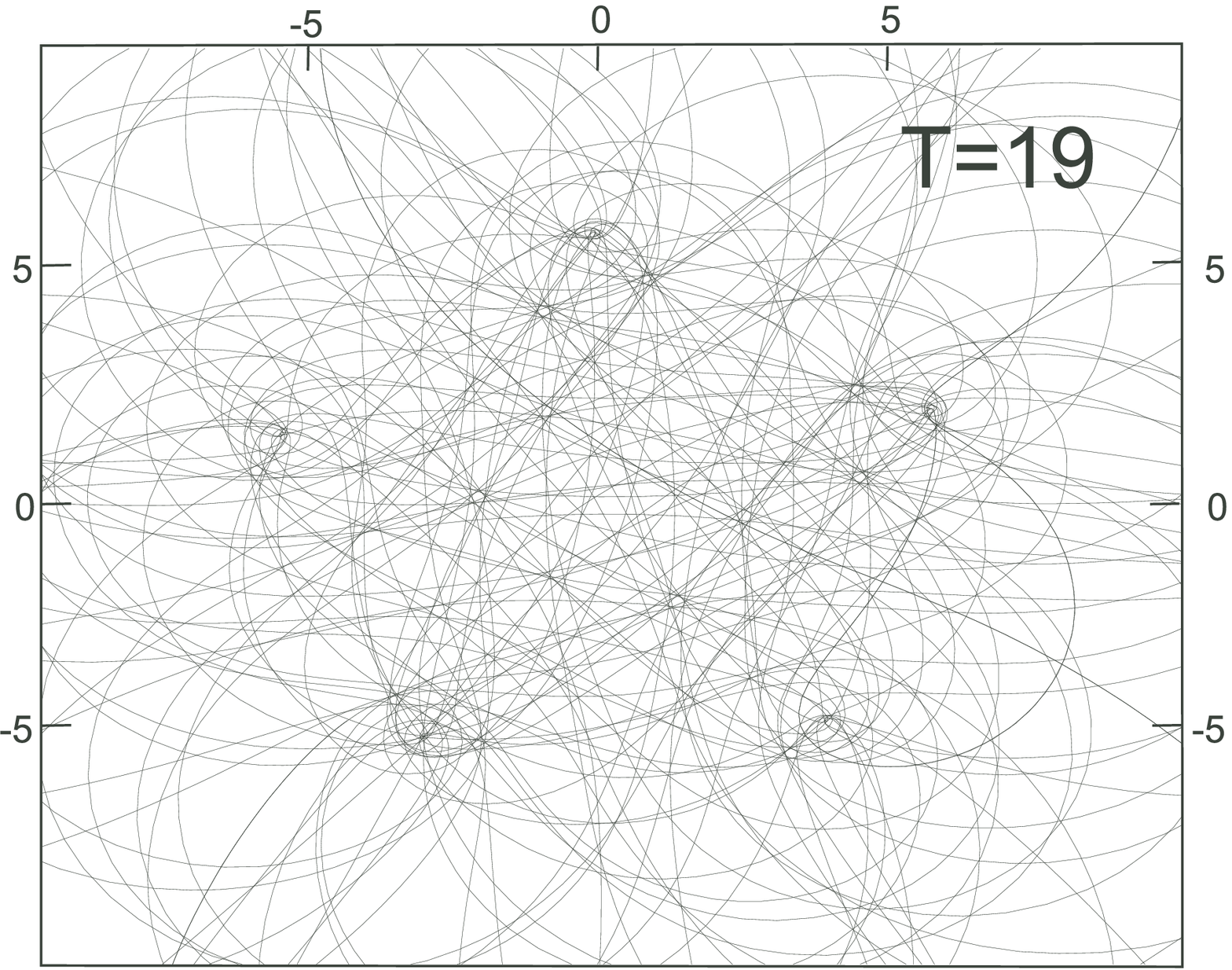,width=2.2 in}
\end{tabular}
\end{tabular}
\end{center}
\caption{Left column: sequences of sheets of $\mathcal R$ visited
every complete turn. Right column: orbits of $\zeta(t)$ obtained
by numerical integration of the system \eqref{eq:zeta(t)} with
$k=4$ for different initial data gathered in Table \ref{tab1}.}
\end{figure}

The path in Figure \ref{fig:fundcell} is discontinuous because
 many of the arcs are contained in other sheets $\mathcal D_{ij}$ of $\mathcal
 R$. Let us consider a sequence of elements of $\mathbb Z^2$
 corresponding to the index of the sheet $\mathcal D_{ij}$ visited
 by the path after every complete turn. For each of the cases
 1A--1D of Table \ref{tab1} this sequence is depicted in the first
 column of Figure \ref{fig:z2}. Note that the sequence in $\mathbb
 Z^2$ is periodic if and only if the orbit of the systems of ODEs
 \eqref{eq:xyk4} is closed.

 Finally, we have performed a numerical integration of the
 equations of motion \eqref{eq:xyk4} for each of the initial data
 in Table \ref{tab1}, using an embedded Runge-Kutta method of order 5(4) with automatic step size control, developed by Prince and Dormand in
 1981. The results of the integration show periodic orbits whose periods are in perfect agreement
 with those predicted by the algorithm of Section
 \ref{sec:circular}. On the right column of Figure \ref{fig:z2} we have displayed the orbits of the function $\zeta(t)$
  whose evolution is described by \eqref{eq:zeta(t)} and is trivially related to
 $z(t) = x(t) + {\rm i} y(t)$ by  \eqref{eq:trick2}.

\subsection{Results for $k=5$}

We study now trajectories of the dynamical system (\ref{eq:xyk5})
corresponding to different initial data. As mentioned in Remark
3.1, this dynamical system describes circular travelling on a
Riemann surface $\mathcal R$ with only two sheets. Therefore
almost all solutions of (\ref{eq:xyk5}) are periodic with period
either $T$ or $2T$. This prediction is confirmed by the numerical
integration of the system and some examples of orbits for
different initial data have been gathered in Table \ref{tab2}.
 In all four cases $2A$--$2D$ the initial velocities $\dot x(0) = \dot y(0) =0$, while the
initial positions $\{x(0),y(0)\}$ are displayed in Table
\ref{tab2}. For simplicity we have also set $\omega=2\pi$ so that
the fundamental period $T=1$.

\begin{table}[h]\caption{Different sets of initial data for $k=5$
\label{tab2}}\vskip 0.2cm
\begin{center}
\begin{tabular}{|c|c|c|c|c|}
  \hline
  Case & Initial data $(x_0,y_0)$ & $|E|$ &  Period \\
  \hline
 2A &$(1.8, 1.9)$ & $3.350 \cdot 10^2$ & $1$ \\
 2B &$(2.8, 2.5)$  & $2.865 \cdot 10^3 $   & $1$ \\
 2C &$(4, 4.5)$  & $4.781\cdot 10^4$ & $1$ \\
 2D &$(5, 8)$  & $7.052\cdot 10^5$  & $1$ \\
  \hline
\end{tabular}
\end{center}
\end{table}

\begin{figure}[h]
\begin{center}
\begin{tabular}[h]{cc}
\begin{tabular}{l}
\psfig{figure=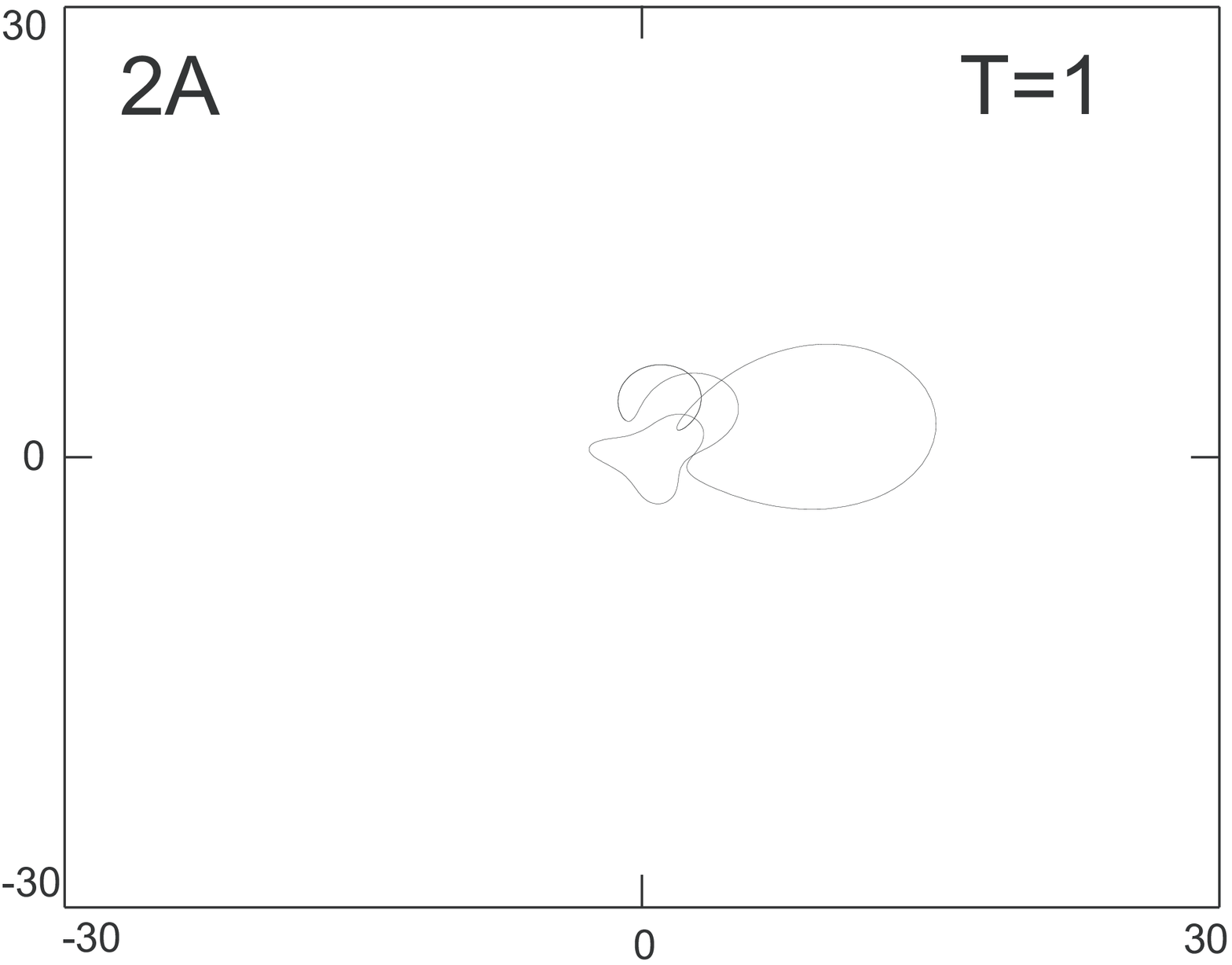,width=2.2in}
\end{tabular} &
\begin{tabular}{l}
\psfig{figure=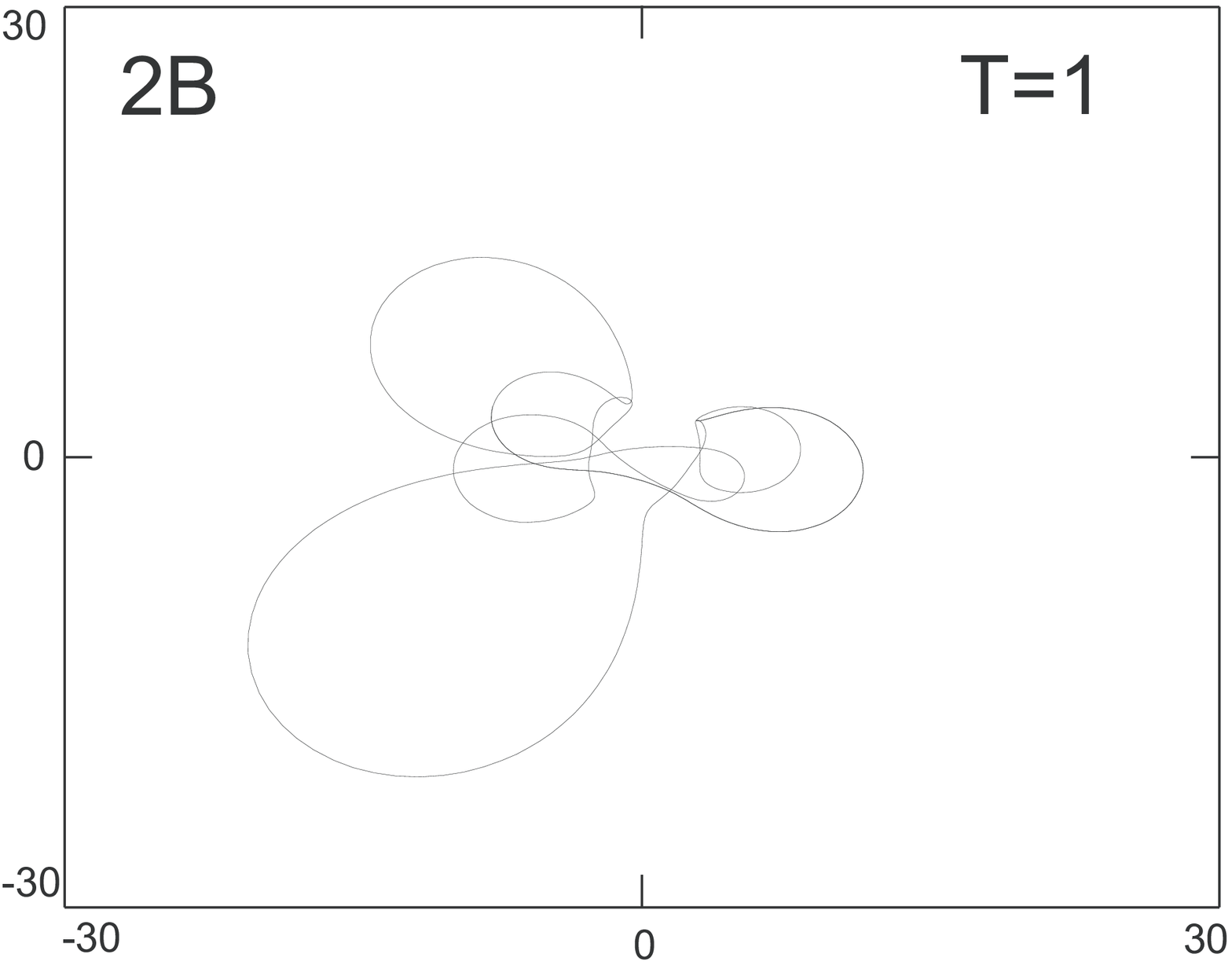,width=2.2in}
\end{tabular}
\\\\
\begin{tabular}{l}
\psfig{figure=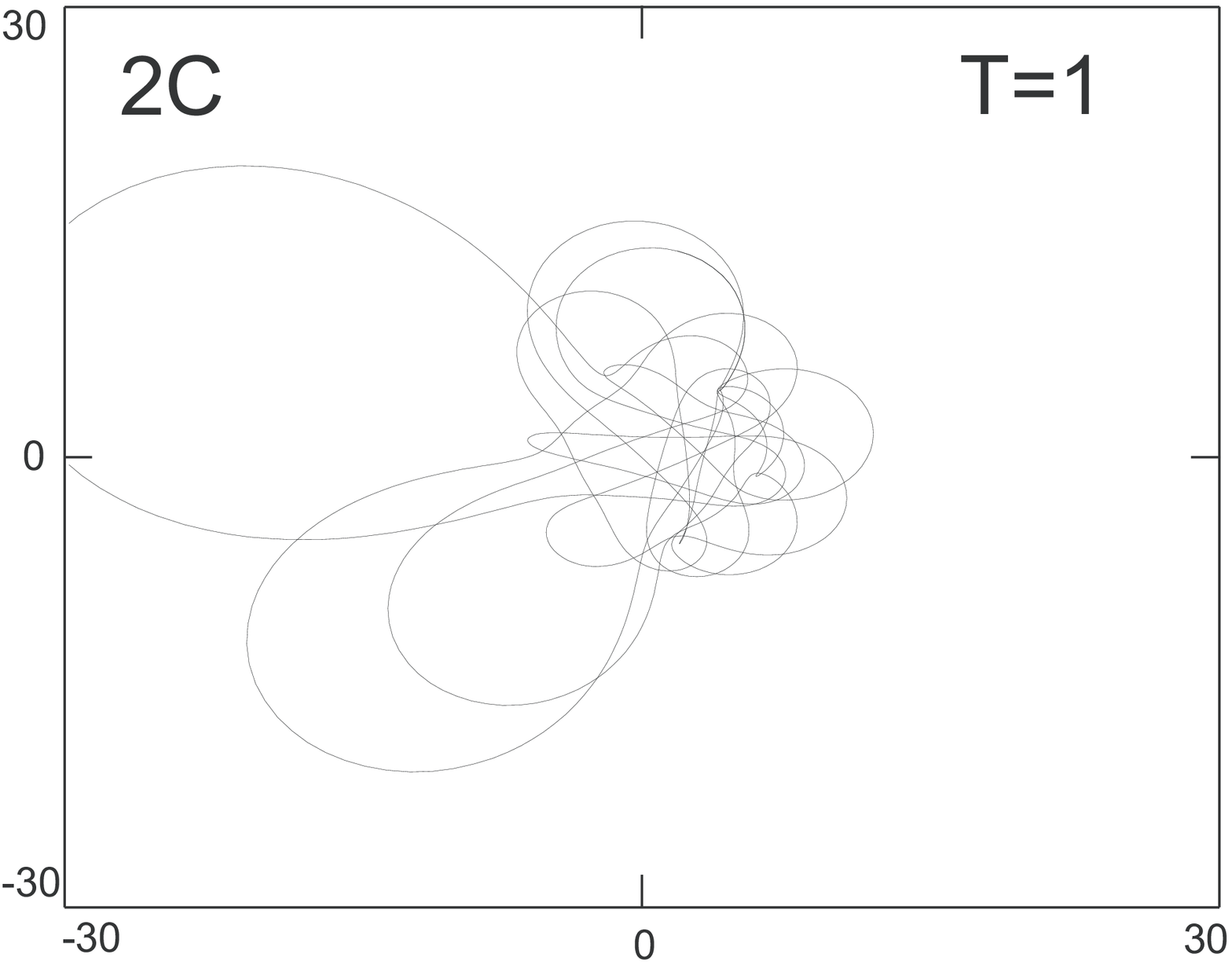,width=2.2in}
\end{tabular}
&
\begin{tabular}{l}
\psfig{figure=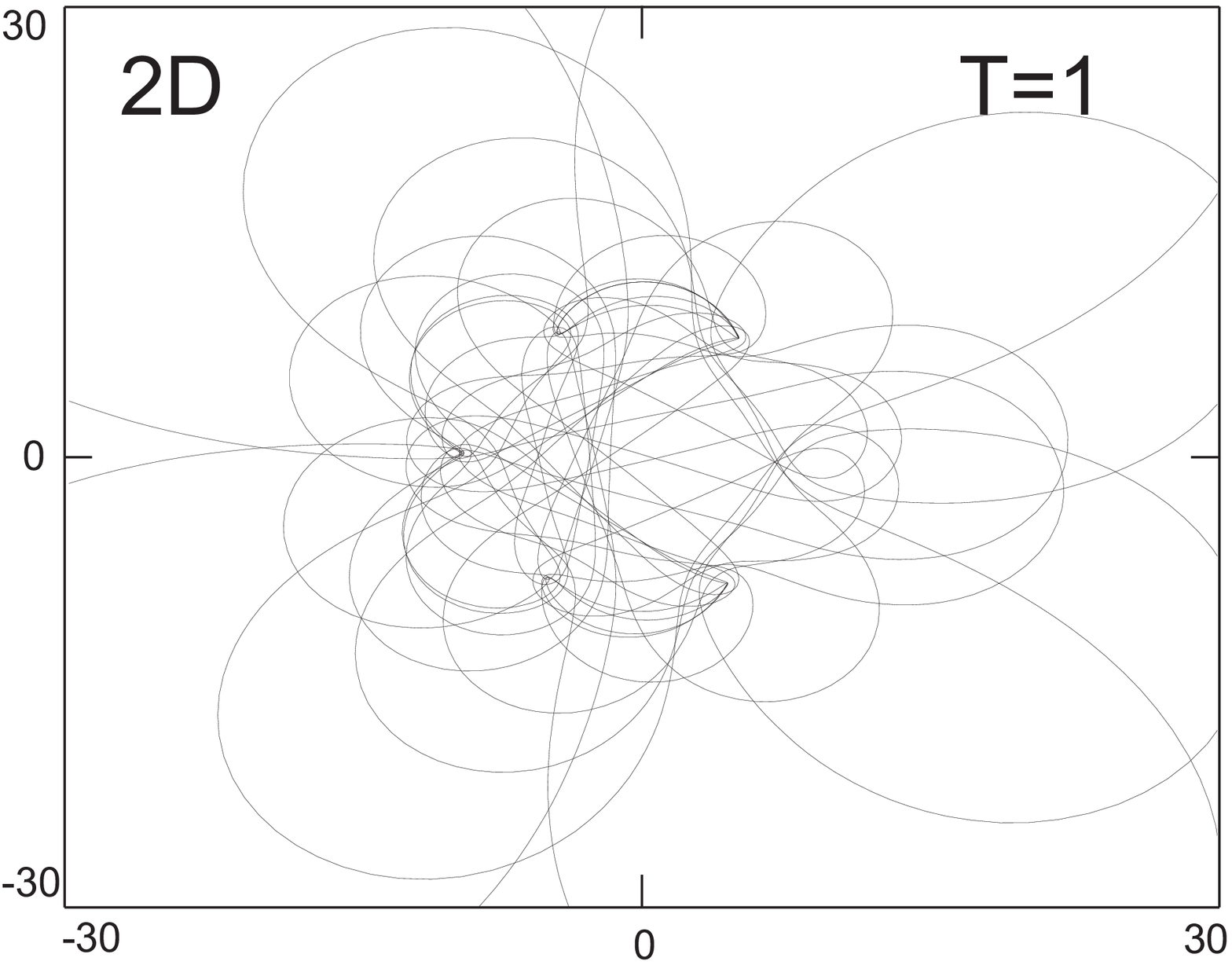,width=2.2in}
\end{tabular}
\end{tabular}
\end{center}
\caption{Orbits of $\zeta(t)$ obtained by numerical integration of
the system \eqref{eq:zeta(t)} with $k=5$ for different initial
data gathered in Table \ref{tab2}.}\label{fig:k51}
\end{figure}

\begin{figure}[h]
\begin{center}
\psfig{figure=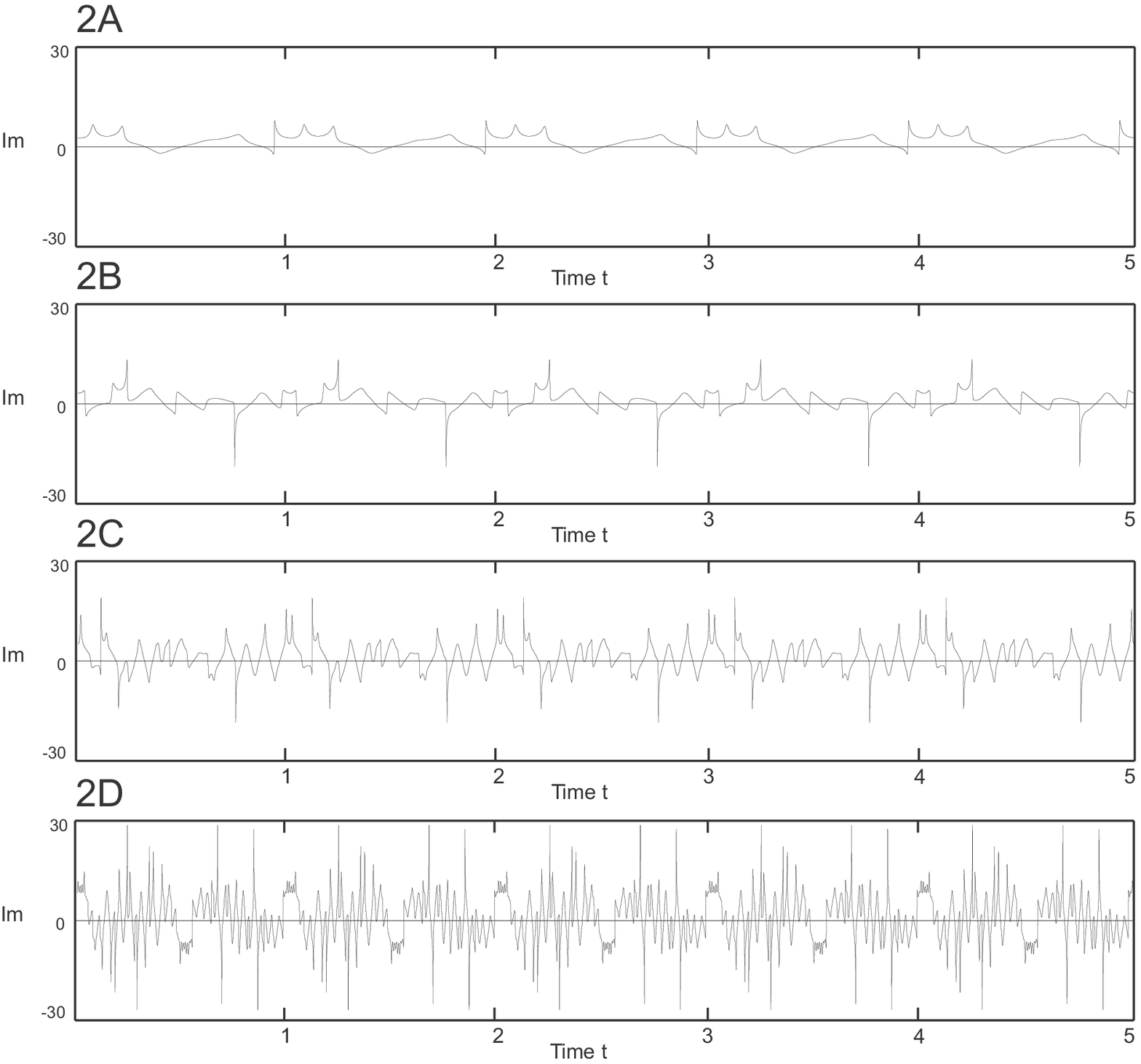,width=5.2in}
\end{center}
\caption{Time evolution of $\textrm{Im}\, \zeta(t)$ obtained by
numerical integration of the system \eqref{eq:zeta(t)} with $k=5$
for different initial data gathered in Table \ref{tab2}. All
solutions are periodic with the same period $T=1$ but their
complexity increases with increasing energy $|E|$.}\label{fig:k52}
\end{figure}

Although all the orbits $2A$--$2D$ have the same period, a look at
Figures \ref{fig:k51} and \ref{fig:k52} reveals that the
complexity of the orbit increases with increasing energy.


\section{Conclusions}\label{sec:conclusions}

In this paper we introduced a class of dynamical systems on
${\mathbb R}^4$ whose solutions can be interpreted as lifting a
circular path on the complex plane to a generally infititely
sheeted Riemann surface $\mathcal R$ describing the inversion of a
hyperelliptic integral. For the integral of the holomorphic
differential $d\eta/\mu$ on a generic genus 2 surface $\Gamma$, we
provided an explicit description of the global structure of
$\mathcal R$  and showed that the surface can be identified with
the universal covering of the theta divisor of Jac($\Gamma$). (It
should be stressed that in case of another holomorphic
differential, in particular $\lambda\, d\lambda/\mu$, the
structure of $\mathcal R$ and type of branching ${\mathcal
R}\mapsto {\mathbb C}$ are different.) We also noted that for some
regular curves with automorphisms, like the sextic curve
$\Gamma_6$, the four periods of the differential are
commensurable, and the corresponding surface $\mathcal R$ becomes
finitely sheeted. This implies that  almost all the trajectories
of the corresponding systems are periodic with period $T$ or $2T$,
although the trajectories increase in complexity as $|E|$
increases.

Using the global description of $\mathcal R$, we provided a
geometric algorithm for the circular travelling on $\mathcal R$,
that replaces the numeric integration of the corresponding ODEs
and associates to the path $\gamma(t)$ on $\mathcal R$  a sequence
on a ${\mathbb Z}^2$-lattice. We conjecture that for a regular
value of the energy, all solutions are periodic, although the
periods can be an arbitrarily high integer multiple of the
fundamental period.

For didactic reasons we have restricted to genus 2 curves
$\Gamma$, although it is also possible to describe the global
geometric structure of Riemann surfaces $\mathcal R$ for
holomorphic integrals on a generic hyperelliptic curve of genus
$g>2$. Namely, dividing its $2g$ periods into two allowed and
$2g-2$ forbidden ones, one constructs an analog of a $\mathcal
D$-sheet by removing from $\mathbb C$ the union of identical
$(4g-4)$-gonal windows $\{{\mathcal W}_{ij}\mid i,j \in {\mathbb
Z}^2\}$ which have pairs of parallel edges formed by the forbidden
periods. Different windows are obtained from each other by shifts
by the allowed periods. Then one can show that the surface
$\mathcal R$ is itself  a union of ${\mathbb Z}^{2g-2}$ identical
copies of $\mathcal D$ glued to each other along the opposite
edges of the windows. A trajectory $\gamma (t)$ on $\mathcal R$
can be encoded as a path on a ${\mathbb Z}^{2g-2}$-lattice.

\subsection*{Acknowledgements} We thank L. Gavrilov, P. Santini, and V. Enolski for
discussions and valuable remarks. Our research was partially
supported by the Spanish Ministry of Science and Technology under
grant BFM 2003-09504-C02-02.


\end{document}